\documentclass[11pt]{article}
\pdfoutput=1

\usepackage[a4paper, total={6in, 8in}]{geometry}
\usepackage{jheppubmod}
\usepackage{mathtools}
\usepackage{psfrag}
\usepackage{array}
\usepackage{amssymb}
\usepackage{amsmath}
\usepackage{cancel}
\usepackage{braket}
\allowdisplaybreaks
\usepackage{mathrsfs}
\usepackage{wrapfig}
\usepackage{xcolor}
\usepackage{physics}
\usepackage{simpler-wick}

\usepackage{amsthm}
\usepackage{graphicx}
\usepackage{caption}
\usepackage{paralist}
\usepackage{comment}
\usepackage[sorting=none]{biblatex}
\newcommand{\secref}[1]{Section \ref{#1}}
\newcommand{\appref}[1]{Appendix \ref{#1}}
\newcommand{\figref}[1]{Fig. \ref{#1}}
\newcommand{\kva}{\ket{0}}
\newcommand{\bva}{\bra{0}}
\newcommand{\kpsi}{\ket{\psi}}
\newcommand{\bpsi}{\bra{\psi}}
\newcommand{\kg}{\ket{g}}
\newcommand{\bg}{\bra{g}}

\title{\textbf{Holography of Information in a Ball of Finite Radius}}

\author{Nava Gaddam$^{\dagger}$ and }
\author{Ashik H$^{\ddag}$} 
\affiliation{$^{\dagger , \ddag}$International Centre for Theoretical Sciences, Tata Institute for Fundamental Sciences, Bengaluru 560089, Karnataka, India.}
\emailAdd{ n.gaddam@icts.res.in}
\emailAdd{ ashik.h@icts.res.in}

\abstract{The principle of holography of information states that all information available in the bulk of asymptotically flat spacetime is also available near its boundary at spatial infinity. However, physical observers never have access to spatial infinity. Therefore, we ask the question: ``Is information contained in a ball of finite radius also holographic in nature?". Phrased differently, we ask whether correlation functions on the boundary of the ball capture all the information of all correlators in the bulk of the ball. In this work, we answer this question in the affirmative within the confines of linearised quantum gravity for generic low-energy (free Klein-Gordon) states in flat space created by the action of unitaries on the vacuum to leading order in a certain perturbative parameter. For an infinite sub-class of these states, our results are perturbatively exact and valid to all orders in this same parameter. Unlike at infinity, massive and massless fields can be captured within the same framework.}
\begin{document}

\date{\today}
\maketitle

\section{Introduction}
There has long been an expectation that gravity is intrinsically non-local, at least far in the ultraviolet. Bekenstein and Hawking's entropy formula for black holes  \cite{Bekenstein:1972tm, Hawking:1975vcx} demonstrates that information is not distributed throughout the volume of space. Instead, information of the deep interior can be seen as constrained to the area of the horizon. This is a hallmark of holography \cite{tHooft:1993dmi, Susskind:1994vu} and indicates that quantum gravity operates with some form of non-local behavior in the ultraviolet regime. Moreover, as we approach very short distances (the Planck scale or the UV limit of quantum gravity), the standard assumptions of locality in quantum field theory break down. The Bekenstein-Hawking entropy scaling reflects that, near the Planck scale, gravitational interactions become strongly coupled, and locality, as defined in ordinary quantum field theories, ceases to be a good approximation. 

Since the advent of gauge/gravity duality \cite{Maldacena:1997re, Gubser:1998bc, Witten:1998qj}, it has become increasingly apparent that the nature of non-locality is not restricted to the ultraviolet regime. In fact, perturbative corrections in the boundary CFT, which reflect quantum gravitational effects in the bulk, introduce non-local behavior even at low energies (in the infrared) through the boundary theory's strongly coupled dynamics. This is manifest in many ways. For instance, graviton exchange in the bulk corresponds to non-local correlations in the boundary operators. Moreover, bulk-to-boundary propagators spread information across large distances since boundary correlators can be influenced by fields in the bulk that are separated by large distances. Finally, $1/N$ corrections introduce inherently non-local quantum corrections which are spread across the boundary theory. 

These ideas warranted a new perspective on the precise holographic nature of information in gravitating theories. A promising picture that has emerged in recent years is the principle of holography of information \cite{Laddha:2020kvp} which states that all information available in the bulk of spacetime is also available near its boundary at spatial infinity. This principle has been formulated in anti-de Sitter \cite{Chowdhury:2020hse}, flat \cite{Chowdhury:2021nxw}, and most recently in De Sitter spaces \cite{Chakraborty:2023los}.  

In all of these developments, the special holographic nature of information in quantum gravity is evident in regimes that are inaccessible to the low-energy bulk observer. One has to go to the ultraviolet regime, or the asymptotic boundary in flat or AdS spaces, or far in the future in de Sitter spacetime. However, physical bulk observers never have access to spatial infinity. Therefore, it begs the question of whether the holographic nature of information in quantum gravity only concerns an intrinsic definition of the theory or if it also has physical relevance for low-energy bulk observers. With this in mind, in this article, we ask the following question: Is information contained in a ball of finite radius also holographic in nature? Phrased differently, we ask whether correlation functions on the boundary of the ball capture all the information of all correlators in the bulk of the ball. Phrased yet another way, we ask how hard it is for information to be hidden in the interior of a ball of finite radius from an observer sitting at its boundary.

A completely general answer to this question is an extremely hard one to find. For example, if there is a black hole (or any sufficiently complicated state) inside the ball of interest, one expects that it is extremely difficult to decode all information about the state from the boundary of the ball (at least with simple measurements). Therefore, as a step in this direction, we will focus on low-energy states in flat space in this article. 

The primary goal of this article is to understand whether, in quantum gravity,  local excitations in a region of space can be detected from far but finite distance away from the said excitations. In quantum field theory, local gauge invariant observables render this impossible. The Gauss law constraint in gravity, however, allows us to measure energy correlations with various boundary operators at infinity which in turn help decode the precise bulk state. In the present article, we will make use of similar energy correlations but now on the boundary of a finite-sized ball to decode the precise local excitations inside the ball. In fact, entaglement between excitations inside and outside the ball allows us to determine some information about the state outside the ball. In order to appreciate precisely what information we might expect to decode, we will first need a quantitative description of the said information inside the ball. To this end, we will now introduce a basis of local excitations.

A natural basis to create such local excitations in a region of space is given by the action of unitary operators built out of all available operators in the theory on the vacuum. We will consider a free massive Klein-Gordon field theory in (3+1)-dimensional Minkowski space. Since we are interested in the space of states created by unitary operators acting on the vacuum, we choose to restrict ourselves to be on the $t=0$ slice in flat space. Then, within the confines of linearised gravity, the most general unitaries are built out of normal ordered products of the Klein-Gordon field and its conjugate momentum as
\begin{equation}\label{eqn:generalUnitary}
    U ~ = ~ \exp\left(i \gamma \sum_{n,m=0}^\infty \int \left(\prod_{i=1}^{n+m} \mathrm{d}^3 \vec{x}_i\right) f_{n,m}\left(\left\{\vec{x}\right\}\right) : \phi^n\left(\left\{\vec{x}\right\}\right) \pi^m\left(\left\{\vec{x}\right\}\right) : \right) \, .
\end{equation}
Here, $f_{n,m}\left(\left\{\vec{x}\right\}\right)$ are multi-local functions that are sufficiently well-behaved at infinity to ensure that the states created by acting with these unitaries on the vacuum approach the vacuum at infinity (namely, they do not change the asymptotic structure of spacetime). The parameter $\gamma$, which we will choose to be small, allows us to keep the excitations controlled to ensure that background field perturbation theory about flat space is not ill-defined. The excitations $U\kva$ would be localised to within a ball of radius $R$ either if we ensure that the functions $f_{n,m}$ have support only inside the ball or if we restrict the integral to be within $\abs{\vec{x}_i} \leq R$. In this article, however, we will be able to derive general results without this restriction and we will allow the unitaries to have support everywhere in space.\footnote{In order to avoid unimportant subtleties with contact terms, we will assume that the state has no support in the region where we make boundary measurements.} 

Given the above excited states that we will consider, namely $\kpsi \coloneqq U\kva$, the next question to ask is for a definition of what we might mean by all information inside the chosen ball of radius $R$. We will define this \textit{information} to be all the knowledge that we may extract about the multi-local functions $f_{n,m}$ from measuring all allowed bulk correlations $\bpsi \pi^m \phi^n \kpsi$ inside the ball. In this article, we will use the following notational legend:
\begin{itemize}
    \item Space-time points that may take values anywhere in space-time will be labelled by $x_i$.
    \item Space-time points that may take values exclusively inside a ball of radius $R$ will be labelled by $y_i$.
    \item Space-time points that may take values exclusively in a small interval near the boundary of the said ball of radius $R$ will be labelled by $z_i$.
\end{itemize}

Therefore, by making measurements of the form $\bpsi \pi^m \left(\left\{\vec{y}\right\}\right) \phi^n \left(\left\{\vec{y}\right\}\right) \kpsi$ for all $\abs{\vec{y}_i} \leq R$, we may invert these observations to obtain partial knowledge of the functions $f_{n,m}$. We do not expect to obtain complete information about these functions as measurements entirely within the ball do not capture all information about the state everywhere on the Cauchy slice.

\paragraph{Results} The primary result of this paper is an explicit demonstration that all information that can be extracted from bulk correlators inside the ball can also be extracted from the boundary\footnote{We allow ourselves a small interval in the radial direction near the boundary.} of the ball, to linear order in $\gamma$ for the states of the form $\kpsi = U \kva$, created by the most general unitary operators available in the theory as defined in \eqref{eqn:generalUnitary}. In other words, we show that bulk and boundary correlators contain the same information about the state. We find that bulk correlators allow us to determine the dependence on the functions $f_n$ on all but one arguments everywhere on the Cauchy slice, while its dependence on one argument is restricted to be within the ball. This knowledge of the functions outside the ball is owed to non-trivial entanglement between the inside and outside of the ball. We find that the same information about the functions can be determined from boundary correlators too. An interesting sub-class of states is when they are created by unitaries which contain either field insertions or conjugate momentum insertions but not both. For this infinite sub-class of states, we find results to all orders in $\gamma$. Our results lend credibility to a notion of holography of information in finite regions of spacetime, without having to go to infinity, at least within the approximations detailed above. We emphasise, again, that such a holographic decoding of the bulk state would be impossible in local quantum field theory and is a feature of quantum gravity.

\paragraph{Organisation of this paper} The boundary correlations that allow us to determine the bulk state involve a certain finite-radius Hamiltionian for the field theory which we describe in \secref{sec:Hsm}, along with a detailed discussion about the states of interest. In \secref{sec:bulkcorrelators}, we derive the bulk correlators in the sub-class of excited states of interest and demonstrate what information they carry about the states. The corresponding boundary calculations are presented in \secref{sec:boundarycorrelators}. We generalise these results to the most general unitary excitations in \secref{sec:generalStates} before concluding with the list of shortcomings of this work in \secref{sec:discussion}. We delegate several technical details and background material to the various appendices at the end of the paper.

\section{Hamiltonian at finite radius and the space of excitations}\label{sec:Hsm}

In this article, we will concern ourselves with a free massive/massless Klein-Gordon field minimally coupled to gravity in asymptotically flat space in $3+1$ dimensions. The Hamiltonian density is given by
\begin{align}\label{eqn:hamiltoniandensity}
    \mathcal{H}\left(t,\vec{x}\right) ~ &\coloneqq ~ \dfrac{1}{2} : \left(\dot{\phi}^2 + \left(\nabla \phi\right)^2 + m^2 \phi^2 \right) : \, ,
\end{align}
Integrating this density over all of space on a given time-slice gives the complete Hamiltonian which is a boundary term at spatial infinity of the said Cauchy slice. We are interested in a finite region of the bulk space and therefore, we will integrate the Hamiltonian density in a finite region of space from, say, $r = 0$ to $r= R$ to define
\begin{align}\label{eqn:HR}
    H \left(R, t\right) ~ &\coloneqq ~ \int_{\left|\vec{x}\right| \leq R} \mathrm{d}^3 \vec{x} \, \mathcal{H}\left(t, \vec{x}\right) ~ = ~ \int_{0}^R r^2 \mathrm{d}r \int_0^\pi \sin\left(\theta\right) \mathrm{d}\theta \int_0^{2 \pi} \mathrm{d} \phi \, \mathcal{H}\left(t, r, \theta, \phi\right) \, ,
\end{align}
This operator is ill-defined owing to large quadratic fluctuations \cite{Bousso:2017xyo, Laddha:2020kvp} in the large-$R$ limit. To tame these, it can be smeared both in time and the radial direction in an appropriate way as has been shown in the case of massless fields in \cite{Laddha:2020kvp}.\footnote{In this reference, it has been shown that there exists a choice of radial smearing that renders quadratic fluctuations of the massless Hamiltonian convergent. A slightly more refined smearing can be employed to tame the divergences for the massive Klein-Gordon case but we will not discuss it in this article as we do not take any large-$R$ limit} We review the technicalities of these issues in \appref{app:Hsmfluctuations}. The smearing in time is governed by a scale which can be tuned to be far in the ultraviolet. Therefore, all measurements that depend on it are sub-leading and the low-energy observer may safely ignore those effects. The radial fluctuations on the other hand, scale with the size of the ball $R$ and go to infinity as $R$ approaches infinity. However, since we will be interested in the information inside a ball of a fixed radius $R$, we will simply sample measurements in a neighbourhood outside the surface. This allows us to essentially work with \eqref{eqn:HR} instead of its smeared counterpart in this paper. Finally, in the presence of such matter sources, Einstein's equations allow us write this finite-region Hamiltonian as a boundary term on the boundary of the ball of interest, see \appref{app:HRboundaryterm}.

\subsection{The space of all excitations}

All states or excitations are defined by the action of insertions of the said scalar field (and its conjugate momenta) on the vacuum. Let us first consider the following class of states
\begin{align}\label{eqn:linearbasis}
    \kpsi ~ = ~ \sum_{n=0}^{n_{\text{max}}} \int \mathrm{d}^3 \vec{x}_1 \dots \mathrm{d}^3 \vec{x}_{n} \, g_n\left(\vec{x}_1 , \dots , \vec{x}_{n}\right) \, :\phi\left(\vec{x}_1\right) \dots \phi\left(\vec{x}_{n}\right): |0\rangle \, ,
\end{align}
where $n_\text{max}$ is a (possibly large) finite number and the field insertions are normal ordered. The term with $n=0$ is to be interpreted as having no integrals or field insertions but merely a constant acting on the vacuum. Let us take two such states, $|\psi_1\rangle$ and $|\psi_2\rangle$ on a Cauchy slice at say $t=0$, that are identical to each other in an open set (which we will call $S$) but may differ from each other outside the said open set $S$. Therefore, this implies that for any operator $\mathcal{O}\left(\vec{y}\right)$ inserted inside the open set $\vec{y} \in S$, we have that it's expectation value is identical in both the states:
\begin{equation}\label{eqn:statesinopenset}
    \langle\psi_1\left| \mathcal{O}\left(\vec{y}\right) \right|\psi_1\rangle ~ = ~ \langle\psi_2\left| \mathcal{O}\left(\vec{y}\right) \right|\psi_2\rangle \, .
\end{equation}
For the states to actually be different outside $S$, they must differ by an operator, say $U$, that has support exclusively outside the open set $|\psi_1 \rangle = U |\psi_2\rangle$. Consider, now, the expectation value of an observable $\mathcal{O}\left(\vec{y}\right)$ inserted inside the open set $\vec{y} \in S$ in the state $|\psi_1\rangle$:
\begin{align}
    \langle\psi_1\left| \mathcal{O}\left(\vec{y}\right) \right|\psi_1\rangle ~ &= ~ \langle\psi_2\left| \mathcal{O}\left(\vec{y}\right) \right|\psi_2\rangle \nonumber \\
    &= ~ \langle\psi_1\left| U^\dagger \mathcal{O}\left(\vec{y}\right) U \right|\psi_1\rangle \nonumber \\
    &= ~ \langle\psi_1\left| \mathcal{O}\left(\vec{y}\right) U^\dagger U \right|\psi_1\rangle \, .
\end{align}
Here, the second equality is owed to the fact that the operator $U$ is supported outside the open set $S$ whereas the operator $\mathcal{O}$ is inserted inside it. Therefore, these are space-like separated and consequently, commuting. Now, for the states to satisfy \eqref{eqn:statesinopenset}, we see that the operator $U$ must be unitary. Since any unitary operator in field theory may be written as\footnote{Of course, in theories with finite dimensional Hilbert spaces, one may have finite-dimensional opeartors that are unitary. In field theory, however, all operators are built out of the fields or their conjugate momenta. The unitary ones are then written as phases in terms of arguments which are functionals of the said field configurations and their conjugate momenta.} $U = e^{i \tilde{U}}$, we have that the state $|\psi_1 \rangle$ can be written as $|\psi_1\rangle = U |\psi_2\rangle = e^{i \tilde{U}} |\psi_2\rangle$. However, this is in contradiction with our original postulate that $|\psi_1\rangle$ may be written as a finite sum \eqref{eqn:linearbasis}. Therefore, we conclude that if two states of the kind \eqref{eqn:linearbasis} are identical in an open neighbourhood, they are necessarily identical everywhere. This implies that making many measurements in any open neighbourhood anywhere in spacetime allows us to determine the state completely. 

Therefore, we learn that the only states for which a notion of holography of information is interesting are those that have superpositions of infinitely many excitations on the vacuum; namely those with $n_{\text{max}} = \infty$ in \eqref{eqn:linearbasis}:
\begin{align}\label{eqn:linearbasis_infinite}
    \kpsi ~ = ~ \sum_{n=0}^\infty \int \mathrm{d}^3 \vec{x}_1 \dots \mathrm{d}^3 \vec{x}_{n} \, g_n\left(\vec{x}_1 , \dots , \vec{x}_{n}\right) \, :\phi\left(\vec{x}_1\right) \dots \phi\left(\vec{x}_{n}\right): |0\rangle \, .
\end{align}

\subsection{A choice of basis of states for all excitations on the vacuum}
We will instead choose a basis in which all states are created by the action of unitary operators on the vacuum. Then, the most general such state in a theory of a single free Klein-Gordon field can be written as
\begin{align}\label{eqn:generalbasisstates}
    |\psi\rangle ~ &= ~ \exp\left( i \gamma \sum_{n,m=0}^\infty \int \left(\prod_{i=1}^{n+m} \mathrm{d}^3 \vec{x}_{i}\right) \, f_{n,m} \left(\left\{\vec{x}\right\}\right) : \phi^n \left(\left\{x\right\}\right) \pi^m \left(\left\{x\right\}\right) : \right) |0\rangle \, ,
\end{align}
where we have used the notation that $\phi^n$ represents \footnote{It is important to stress that the notation is meant to specify how many arguments there are in the set and not necessarily what those arguments are. For instance, we may also have that $\phi^2 \left(\left\{\vec{x}\right\}\right) = \phi\left(\vec{x}_3\right) \phi\left(\vec{x}_4\right)$.}
\begin{equation}
    \phi^n \left(\left\{x\right\}\right) ~ \coloneqq ~ \phi\left(\vec{x}_1\right) \dots \phi\left(\vec{x}_n\right) \, .
\end{equation}
Here, $\gamma$ is a freely tune-able parameter which we may choose to be small if need be. For the case of either $n=0$ or $m=0$, we will work to all orders in $\gamma$. For the general states where neither is vanishing, we will present our arguments to first order in $\gamma$. Furthermore, the (real) functions $f_{n,m}$ encode all the information of where the states are supported in spacetime. Restricting their domain of dependence to be within the ball will generate completely localised states inside the ball. Writing the scalar field operator in its momentum modes and expanding the exponential, it is easy to check that any state can be written in the above form. 

In this article, we will consider general states which may be supported everywhere in spacetime. However, we will assume that there exists some region of spacetime (at some finite radius $R$ from the origin) where the functions have no support.\footnote{This is to ensure that we may setup our boundary measurements at this radius $R$ without the effect of contact terms. Of course, we may not know \textit{a priori} where to set up our boundary measurements. However, we may use the expectation value of the finite-radius Hamiltonian to identify such a region. The quantity $\bpsi H\left(R,0\right)\kpsi$ measures the energy inside the ball of radius $R$ on the $t=0$ slice as we demonstrate in \eqref{eqn:HRasenergy}. Therefore, by varying the radius at which we make these energy measurements, we may identify a radius at which the energy is below our experimental sensitivity and ensure that the variation of the energy in spheres of nearby radii do not vary too much.} Assuming such a region exists, we proceed to choose our ball of interest where we make holographic measurements to be at this radius. This choice also ensures the vanishing of the expectation value of any (composite) observable\footnote{By a composite observable, we just mean a product of local observables.} that takes the following form: 
\begin{align}
    \bpsi \mathcal{O}\left(\vec{z}_0 , \left\{\vec{z}_i\right\}\right) \kpsi ~ &= ~ \int \left(\prod_{i=1}^{n+m} \mathrm{d}^3 \vec{x}_i\right) f_{n,m} \left( \left\{\vec{x}\right\}\right) K\left(\left\{\vec{x}\right\} , \left\{\vec{z}_i\right\}\right) \delta\left(\vec{x}_0 - \vec{z}_0\right) \qquad \nonumber \\
    &\sim ~ 0 \, .
\end{align}
This is because the Dirac-delta function clicks at the location of the measurement where, by assumption, the functions defining the state $f_{n,m}$ have no support.

\subsection{A sub-space of unitary excitations on the vacuum}\label{sec:states}
An interesting sub-class of the most general states defined in \eqref{eqn:generalbasisstates} arises when the defining functions $f_{n,m}\left(\left\{\vec{x}\right\}\right)$ are taken to vanish for all $m\neq 0$, namely $f_{n,m}\left(\left\{\vec{x}\right\}\right) = f_n\left(\left\{\vec{x}\right\}\right)$. Here, the functions $f_n\left(\left\{\vec{x}\right\}\right)$ are also real and the corresponding states still have unit norm. Such states are then written as:
\begin{align}\label{eqn:basisState}
    |\psi\rangle ~ &\coloneqq ~ e^{-X} \kva ~ = ~ \exp\left( i \gamma \sum_{n=0}^\infty \int \left(\prod_{i=1}^n \mathrm{d}^3 \vec{x}_i\right) \, f_n \left(\left\{\vec{x}\right\}\right) : \phi^n \left(\left\{x\right\}\right) : \right) |0\rangle \, ,
\end{align}
where the functions $f_n$ are assumed to be real, smooth and continuous, $x_i$ label the $n$ points of scalar field insertions, and $\gamma$ is the same small parameter that generates a Madhavan-Taylor series of perturbative excitations above the vacuum. A natural way to consider localised states inside the ball of radius $R$ is to restrict the functions $f_n$ to have support within the ball or to simply restrict the integrals as
\begin{align}\label{eqn:locstate}
    |\psi\rangle ~ &= ~ \exp\left( i \gamma \sum_{n=0}^\infty \int_{\left|\vec{x}\right| < R} \left(\prod_{i=1}^n \mathrm{d}^3 \vec{x}_i\right) \, f_n \left(\left\{\vec{x}\right\}\right) : \phi^n \left(\left\{x\right\}\right) : \right) |0\rangle \, .
\end{align}

\section{Information inside a ball of finite radius}\label{sec:bulkcorrelators}
\begin{figure}[h!]
    \centering
    \includegraphics[scale=0.3]{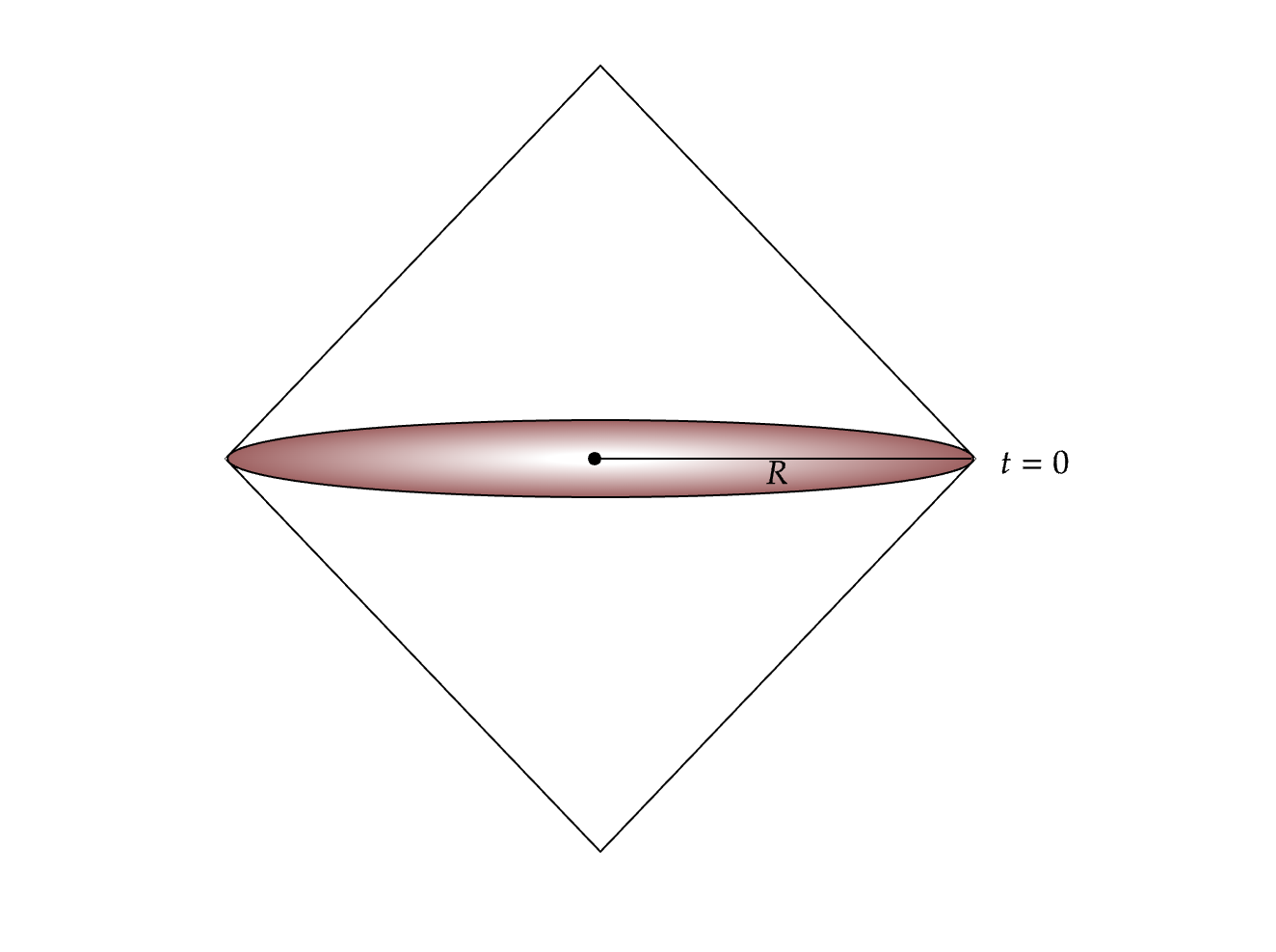}
    \caption{Causal diamond of a ball of radius $R$. We take all information inside the ball to be defined by the set of all correlators on the $t=0$ slice inside the ball.}
    \label{fig:causalDiamond}
\end{figure}

In order to have a notion of holography of information at finite distances, we first need a quantitative notion of all the information contained in the finite sized region of interest. At some time slice, say $t=0$, all the information inside a ball of some finite radius is quantified by all the correlators in the bulk of the ball. Since we are working in linearised gravity minimally coupled to a scalar field theory about flat space, this includes all $n$-point functions built out of the scalar field and its conjugate momentum of the form $\langle \phi\left(\vec{y}_1\right) \phi\left(\vec{y}_2\right) \dots \pi\left(\vec{y}_i\right) \pi\left(\vec{y}_{i+1} \right) \dots \rangle$ where $\left|\vec{y}_j\right| < R ~ \forall ~ j$ lie inside the ball on the same time slice with $y^0_j = 0$, see \figref{fig:causalDiamond}.\footnote{In field theory on a fixed background, causality ensures that these correlators contain all information in the causal diamond of the ball.} To compute the correlators in an arbitrary state, we will need the various vacuum correlators in the theory listed in \appref{app:bulkcorrelators}.

Given the choice of basis of states that we have made, it turns out that certain class of correlators are more convenient to work with. This class of correlators that we will initially attend to are of the form $\bpsi \pi \phi^n\kpsi$ for all $n\geq 0$. There is nothing special about this class. Had we chosen a different basis of states (built out of the conjugate momentum instead of the scalar field itself for example), a different class of correlators would turn out to be convenient to begin with (of the kind $\bpsi \phi \pi^n\kpsi$ for instance). The reason for this convenience will soon become apparent. At the outset, with our choice of basis for the states as in \eqref{eqn:basisState}, we first consider all correlators of the following kind:
\begin{align}
    \bpsi Y \kpsi ~ &\eqqcolon ~ \bpsi \phi\left(y_1\right) \dots \phi\left(y_n\right)\kpsi \nonumber \\
    &= ~ \bva e^X Y e^{-X} \kva \nonumber \\
    &= ~ \bva Y \kva \, ,
\end{align}
where we used the Baker-Campbell-Hausdorff formula: 
\begin{align}\label{eqn:BCH}
    e^X Y e^{-X} ~ = ~ Y + \left[X, Y\right] + \dfrac{1}{2} \left[X , \left[X, Y\right]\right] + \dots 
\end{align}
These correlators reduce to vacuum correlators because the state and all insertions are defined on the same time-slice and all commutators of the field with itself at spacelike separated points are identically vanishing. Therefore, in addition to the correlators of the form $\bpsi \pi \phi^n\kpsi$ that we will begin with, the only other correlators of interest take the form $\bpsi \pi^m \phi^n\kpsi$ where $m>1$ and $n\geq 0$. 

\subsection{General bulk correlators in a sub-space of all unitary excitations}

In what follows, we will look at the difference between vaccum correlators and the correlators in the excited state:
\begin{align}\label{eqn:bulk-correlators}
    &\bpsi \pi\left(\vec{y}\right) \phi^m\left(\left\{\vec{y}_m\right\}\right) \kpsi - \bva \pi\left(\vec{y}\right) \phi^m\left(\left\{\vec{y}_m\right\}\right) \kva ~ = ~ \nonumber \\
    &\quad = - i \gamma \int \sum_{p=1}^\infty \left(\prod_{n=1}^p \mathrm{d}^3 \vec{x}_n\right) f_p\left(\left\{\vec{x}\right\}\right) \Bigg[ \bva \left[:\phi^p\left(\left\{\vec{x}\right\}\right): , \pi(\vec{y})\right] \phi^m\left(\left\{\vec{y}\right\}\right)\kva \nonumber \\
    &\hspace{5cm} + \bva\pi\left(\vec{y}\right) \underbrace{\left[\phi^p\left(\left\{\vec{x}\right\}\right) , \phi^m\left(\left\{\vec{y}\right\}\right) \right]}_{= ~ 0} \kva \Bigg] + \mathcal{O}\left(\gamma^2\right) \nonumber \\
    &\quad = - i \gamma \int \sum_{i = 1}^\infty \left(\prod_{j=1}^{i} \mathrm{d}^3 \vec{x}_j\right) f_i\left(\left\{\vec{x}\right\}\right) \sum_{k=1}^i \bva \left[\phi\left(\vec{x}_k\right) , \pi\left(\vec{y}\right)\right] :\phi^{i-1}\left(\left\{\vec{x}\right\}\right): \phi^m\left(\left\{\vec{y}\right\}\right) \kva \nonumber \\
    &\hspace{5cm} + \mathcal{O}\left(\gamma^2\right) \quad \text{where } \phi^{i-1}\left(\left\{\vec{x}\right\}\right) \text{has no insertion at } \vec{x}_k \nonumber \\
    &\quad = - i \gamma \int \sum_{i = 1}^\infty \left(\prod_{j=1}^{i} \mathrm{d}^3 \vec{x}_j\right) f_i\left(\left\{\vec{x}_i\right\}\right) \sum_{k=1}^i \bva :\phi^{i-1}\left(\left\{\vec{x}_{i-1}\right\}\right): \phi^m\left(\left\{\vec{y}\right\}\right) \kva \left[i \delta^3\left(\abs{\vec{x}_k - \vec{y}}\right)\right] \nonumber \\
    &\hspace{7cm} \text{where } \phi^{i-1}\left(\left\{\vec{x}\right\}\right) \text{has no insertion at } \vec{x}_k \, .
\end{align}
Since the normal-ordered insertions cannot Wick-contract among themselves, it is evident that this correlator is only non-vanishing for the case when $m < i - 1$. This has the rather nice consequence that the correlator of the form $\langle \pi \phi^m\rangle$ isolates the contribution of only terms up to $f_{m+1}$. Therefore, starting with the lowest values of $m$, we may iteratively determine all contributions. Furthermore, we have dropped all higher order terms because they are all identically zero for the following reason. The terms of order $\mathcal{O}\left(\gamma^2\right)$ in the above expression are given by a commutation of the field insertions arising from the state with the $\mathcal{O}\left(\gamma\right)$ term as can be seen from \eqref{eqn:BCH}. Since the latter is entirely given in terms of the field and there are no remnant conjugate momentum insertions left, these commutators are all identically vanishing when the insertions are all at space-like separated points. The same conclusion holds for terms of all higher orders. This is the reason for beginning with correlators of the form $\bpsi \pi \phi^n\kpsi$; the linearised expansion is exact. It is now also clear that all other correlators of the form $\bpsi \pi^m \phi^n\kpsi$ will terminate at order $\mathcal{O}\left(\gamma^m\right)$.

\subsection{One-point functions}\label{sec:bulkonept}
There are two one-point functions available, $\bpsi \phi\left(y\right) \kpsi$ and $\bpsi \pi\left(y\right) \kpsi$. The former is vanishing at all orders of $\gamma$. The one-point function of the conjugate momentum $Y \coloneqq \pi \left(y\right)$ is obtained by specifying \eqref{eqn:bulk-correlators} to the case of $m=0$. In this case, since the insertions coming from the state are normal-ordered, contractions between themselves are vanishing. Therefore, only the one-particle state contributes to this correlation function resulting in
\begin{align}\label{eqn:one-particle-info}
    \bpsi \pi\left(y\right) \kpsi ~ &= ~ \gamma \int \mathrm{d}^3 \vec{x} ~ f_1 \left(\vec{x}\right) \delta^3\left(\abs{\vec{x} - \vec{y}}\right) ~ = ~ \gamma f_1\left(\vec{y}\right) \, .
\end{align}

\subsection{Two-point functions}
Following the previous calculation, it is straightforward to check that the two-point functions isolate the two-particle contribution, again owing to the normal ordering in the state. We find the following for the $\langle\pi \phi\rangle$ correlator is
\begin{align}\label{eqn:bulktwoptfn}
    &\bpsi \pi\left(\vec{y}_1\right) \phi\left(\vec{y}_2\right) \kpsi - \bva \pi\left(\vec{y}_1\right) \phi\left(\vec{y}_2\right) \kva ~ = ~ \nonumber \\
    &\quad = ~ - i \gamma \int \mathrm{d}^3 \vec{x}_1 \mathrm{d}^3 \vec{x}_2 f_2(\vec{x}_1 , \vec{x}_2) \nonumber \\
    &\hspace{2cm} \times \left(\dfrac{i m \delta^3 \left(\vec{x}_1 - \vec{y}_1\right)}{4 \pi^2 \left|\vec{y}_2 - \vec{x}_2\right|} K_1\left(m \left|\vec{y}_2 - \vec{x}_2\right|\right) + \dfrac{i m \delta^3 \left(\vec{x}_2 - \vec{y}_1\right)}{4 \pi^2 \left|\vec{y}_2 - \vec{x}_1\right|} K_1\left(m \left|\vec{y}_2 - \vec{x}_1\right|\right) \right) \nonumber \\
    &\quad = ~ \dfrac{m \gamma}{4 \pi^2} \int \mathrm{d}^3 \vec{x}_1 \left(\dfrac{K_1\left(m \left|\vec{y}_2 - \vec{x}_1\right|\right) }{ \left|\vec{y}_2 - \vec{x}_1\right|}\right) \Big[f_2\left(\vec{x}_1 , \vec{y}_1\right) + f_2\left(\vec{y}_1 , \vec{x}_1\right)\Big] \nonumber \\
    &\qquad \eqqcolon ~ \dfrac{m \gamma}{4 \pi^2} \int \mathrm{d}^3 \vec{x}_1 \left(\dfrac{K_1\left(m \left|\vec{y}_2 - \vec{x}_1\right|\right) }{ \left|\vec{y}_2 - \vec{x}_1\right|}\right) \tilde{f}_2\left(\vec{x}_1 , \vec{y}_1\right) \, ,
\end{align}
which includes the standard two-point function that we derive in \appref{app:bulkcorrelators-massive}. It is now interesting to ask what knowledge of the state inside the ball means. First, we note that two states with the arguments of the function $f\left(x_1 , x_2\right)$ switched cannot be distinguished; this explains the appearance of the symmetric combination. Since $y_i$ are meant to be insertions inside the ball of size $R$, one would expect that this correlator determines the function $\tilde{f}_2\left(\vec{y}_1 , \vec{y}_2\right)$ completely inside the ball. 

In \appref{app:bulkinversion}, we demonstrate exactly what information we may infer about this function given the above correlators. By definition, we take this to define the information about the two-particle state that is contained in the ball. As we demonstrate in the appendix and argue below \eqref{eqn:inversionbulkf2}, it is possible to invert the above equation \eqref{eqn:bulktwoptfn}. What we find is that the dependence of the function $\tilde{f}_2\left(\vec{x}_1 , \vec{x}_2\right)$ on one of the arguments, $\vec{y}_1$, can be reconstructed entirely inside the ball. Whereas the dependence on the second argument can be reconstructed globally, everywhere in space! Since the states $\psi$ in our basis of choice are built out of field insertions, the insertion of the conjugate momentum in the correlators $\pi\left(\vec{y}_1\right)$ allows us to measure its expectation value in the presence of the other field insertion $\phi\left(\vec{y}_2\right)$ which is entangled with the field insertions $\phi\left(\vec{x}\right)$, arising from the state, that may be distributed everywhere in spacetime. Therefore, the conjugate momentum insertion (which is confined to be within the bulk of the ball by construction) limits our knowledge of $\tilde{f}_2\left(\vec{y}_1 , \vec{x}_2\right)$ in one argument to within the ball whereas the entanglement allows us to reconstruct its dependence on the second argument everywhere in space. This defines for us, what information about the state we may decode by making bulk measurements.

\subsection{Three-point, higher-point, and general correlators}
In similar vein, further multi-particle contributions can be calculated. For instance, the three-particle contribution is found to be
\begin{align}
    &\bpsi \pi\left(\vec{y}_1\right) \phi\left(\vec{y}_2\right) \phi\left(\vec{y}_3\right)\kpsi ~ = ~ \nonumber \\
    & = ~ i \gamma \int \left(\prod_{j=1}^{3} \mathrm{d}^3 \vec{x}_j\right) f_3\left(\vec{x}_1, \vec{x}_2 , \vec{x}_3\right) \sum_{k=1}^i \bva :\phi^{i-1}\left(\left\{\vec{x}\right\}\right): \phi^m\left(\left\{\vec{y}\right\}\right) \kva \left[i \delta^3\left(\abs{\vec{x}_k - \vec{y}}\right)\right] \nonumber \\
    &\hspace{3cm} + \text{terms containing } f_1 \nonumber \\
    &= ~ - \gamma \int \left(\prod_{i = 1}^3 \mathrm{d}^3 \vec{x}_i\right) f_3 \left(\vec{x}_1 , \vec{x}_2, \vec{x}_3\right) \bva \Bigg[\delta^3\left(\vec{x}_3 - \vec{y}_1\right) \phi\left(\vec{x}_1\right) \phi\left(\vec{x}_2\right) \phi\left(\vec{y}_2\right) \phi\left(\vec{y}_3\right) \nonumber \\
    &\hspace{6.3cm} + \delta^3\left(\vec{x}_2 - \vec{y}_1\right) \phi(\vec{x}_1) \phi(\vec{x}_3) \phi(\vec{y}_2) \phi(\vec{y}_3) \nonumber \\
    &\hspace{6.3cm} + \delta^3\left(\vec{x}_1 - \vec{y}_1\right) \phi\left(\vec{x}_2\right) \phi\left(\vec{x}_3\right) \phi\left(\vec{y}_2\right) \phi\left(\vec{y}_3\right) \Bigg] \kva \nonumber \\
    &\hspace{3cm} + \text{terms containing } f_1 \, .
\end{align}
Having determined $f_1$ already, following the algorithm presented in \appref{app:bulkinversion}, we may also invert this integral equation by making multiple measurements for various values of $\vec{y}_i$, to find the function $\tilde{f}_i\left(\vec{y}_1 , \left\{\vec{x}_{i-1}\right\}\right)$. Generically such higher-point functions involve the information of the support functions $f_j$ with $j<i$ that correspond to lower-point functions. Since we have already determined those, we may use that knowledge to invert for the higher-point functions. As we argued in the preceding sub-section, we will have knowledge of the function inside the ball in one argument $\vec{y}_1$ whereas we can reconstruct its dependence on all other arguments explicitly everywhere in space. 

So far, we have focused our attention on correlators of the form $\langle \pi \phi^m\rangle$. This was because, given our choice of basis of states, such correlators conveniently isolated the functions defining the state. Using many such measurements, we were able to completely reconstruct these functions that define the state inside the ball (and partially outside it). There are of course many more correlators available in the theory: $\langle\pi^m \phi^n\rangle$. These correlators, however, generically contain information of all the functions $\tilde{f}_i\left(\vec{y}_1 , \left\{\vec{x}_{i-1}\right\}\right)$ which we have already found. Given our knowledge of these functions, we may simply compute these correlators explicitly thereby gaining all information of all bulk correlators inside the ball.

\section{Determining information inside the ball from its boundary}\label{sec:boundarycorrelators}

The primary question of interest is whether we may make observations on the boundary of the ball of interest that allow us to determine all the information of the state inside the ball. This amounts to determining all the bulk correlators that we computed in \secref{sec:bulkcorrelators}. Much like we did in the bulk of the ball, we will consider a particular class of boundary observables to begin with, namely $\langle H_{\text{sm}}\pi \phi^r\rangle$, where $H_{\text{sm}}$ is the smeared Hamiltonian that we defined in \secref{sec:Hsm} and the field insertions are all on the boundary/surface of the ball at the same $t=0$ time slice as in the bulk.

\subsection{General boundary correlators in a sub-space of all unitary excitations}
For the unitary excitations considered in the previous section, we have
\begin{align}\label{eqn:boundary-correlators1}
    &\bpsi H_{\text{sm}} \pi\left(\vec{z}\right) \phi^r\left(\left\{\vec{z}\right\}\right) \kpsi - \bva H_{\text{sm}} \pi\left(\vec{z}\right) \phi^r\left(\left\{\vec{z}\right\}\right) \kva ~ = ~ \nonumber \\
    &\quad = - i \gamma \int \sum_{p=1}^\infty \left(\prod_{n=1}^p \mathrm{d}^3 \vec{x}_n\right) f_p\left(\left\{\vec{x}\right\}\right) \Bigg[ \bva \left[:\phi^n\left(\left\{\vec{x}\right\}\right): , H_{\text{sm}}\right] \pi\left(\vec{z}\right) \phi^r\left(\left\{\vec{z}\right\}\right)\kva \nonumber \\
    &\hspace{5cm} + \bva H_{\text{sm}} \left[\phi^n\left(\left\{\vec{x}\right\}\right) , \pi\left(\vec{z}\right) \right] \phi^r\left(\left\{\vec{z}\right\}\right) \kva\nonumber \\
    &\hspace{5cm} + \bva H_{\text{sm}} \pi\left(\vec{z}\right) \underbrace{\left[\phi^n\left(\left\{\vec{x}\right\}\right) , \phi^r\left(\left\{\vec{z}\right\}\right) \right]}_{= ~ 0} \kva \Bigg] + \mathcal{O}\left(\gamma^2\right) \nonumber \\
    &\quad = - i \gamma \int \sum_{i = 1}^\infty \left(\prod_{j=1}^{i} \mathrm{d}^3 \vec{x}_j\right) f_i\left(\left\{\vec{x}\right\}\right) \nonumber \\ 
    &\hspace{2cm} \times \sum_{k=1}^i \bva \Bigg[:\phi^{i-1}\left(\left\{\vec{x}\right\}\right): \left[\phi\left(\vec{x}_k\right) , H_{\text{sm}}\right] \pi\left(\vec{z}\right) \phi^r\left(\left\{\vec{z}\right\}\right) \nonumber \\
    &\hspace{3cm} + H_{\text{sm}} :\phi^{i-1}\left(\left\{\vec{x}\right\}\right): \left[\phi\left(\vec{x}_k\right) , \pi\left(\vec{z}\right)\right] \phi^r\left(\left\{\vec{z}\right\}\right) \Bigg] \kva + \mathcal{O}\left(\gamma^2\right) \, .
\end{align} 
The smeared Hamiltonian inserted here is smeared in time as defined in \eqref{eqn:Hsm}. This smearing was to ensure that quadratic fluctuations are finite. However, the smearing parameter $\eta$ results in a very sharply peaked contribution around the $t=0$ time slice. All sub-leading effects in $\eta$ are essentially ultraviolet effects. Therefore, for the purpose of low-energy observations of the kind we are interested in, we may take the $\eta \rightarrow 0$ limit which dictates that we replace $H_{\text{sm}}$ by the finite-radius Hamiltonian $H_R \coloneqq H\left(R, 0\right)$ from the definition in equation \eqref{eqn:HR}. Therefore, we have for the above correlator that
\begin{align}\label{eqn:boundary-correlators2}
    &\bpsi H_R \pi\left(\vec{z}\right) \phi^r\left(\left\{\vec{z}\right\}\right) \kpsi - \bva H_R \pi\left(\vec{z}\right) \phi^r\left(\left\{\vec{z}\right\}\right) \kva ~ = ~ \nonumber \\
    &\quad = ~ - i \gamma \int \sum_{i = 1}^\infty \left(\prod_{j=1}^{i} \mathrm{d}^3 \vec{x}_j\right) f_i\left(\left\{\vec{x}\right\}\right) \nonumber \\
    &\hspace{1.5cm} \times \sum_{k=1}^i \Bigg[\bva H_R :\phi^{i-1}\left(\left\{\vec{x}\right\}\right): \phi^r\left(\left\{\vec{z}\right\}\right) \kva \left[i \delta^3\left(\abs{\vec{x}_k - \vec{z}}\right)\right] \nonumber \\
    &\hspace{1.75cm} + i \bva :\phi^{i-1}\left(\left\{\vec{x}\right\}\right): \pi\left(\vec{x}_k\right) \pi\left(\vec{z}\right) \phi^r\left(\left\{\vec{z}\right\}\right) \kva \Theta\left( R - \abs{\vec{x}_k}\right) \Bigg] + \mathcal{O}\left(\gamma^2\right) \, ,
\end{align}
where we used that 
\begin{align}
    \lim_{\eta\rightarrow 0} \left[H_{\text{sm}} , \phi(0,\vec{x})\right] ~ &= ~ - i \pi(\vec{x}) \Theta\left( R - \abs{\vec{x}}\right) \, .
\end{align}
We will first focus our attention on the first term containing the the Dirac-delta function. If the state is completely localised to have support inside the ball, the delta function never clicks and we would only be left with the second term. If the state generically has support everywhere, on the other hand, we will now argue that there is a local measurement that we can do to subtract this contribution. Let us consider the following observable:
\begin{align}
    &\bpsi \pi\left(\vec{z}\right) \phi^{r+1}\left(\left\{\vec{z}\right\}\right) \kpsi - \bva \pi\left(\vec{z}\right) \phi^{r+1}\left(\left\{\vec{z}\right\}\right) \kva ~ = ~ \hspace{6.5cm}\nonumber \\
    &= ~ - i \gamma \int \sum_{i = 1}^\infty \left(\prod_{j=1}^{i} \mathrm{d}^3 \vec{x}_j\right) f_i\left(\left\{\vec{x}\right\}\right) \sum_{k=1}^i \bva :\phi^{i}\left(\left\{\vec{x}\right\}\right): \phi^r\left(\left\{\vec{z}\right\}\right) \kva \left[i \delta^3\left(\abs{\vec{x}_k - \vec{z}}\right)\right] .
\end{align}
This is identical to a generic bulk observable of the kind we calculated in \secref{sec:bulkcorrelators}. From such a correlator, we may infer the functions $f_n\left(\left\{\vec{x}\right\}\right)$ where one of the arguments is fixed to be the bulk point $\vec{z}$ as we demonstrate in \appref{app:bulkinversion}. Therefore, we may plug the resulting knowledge of the functions into the term containing the Dirac-delta in \eqref{eqn:boundary-correlators1}:
\begin{align}\label{eqn:Z}
    Z ~ &\coloneqq ~ - i \gamma \int \sum_{i = 1}^\infty \left(\prod_{j=1}^{i} \mathrm{d}^3 \vec{x}_j\right) f_i\left(\left\{\vec{x}\right\}\right) \nonumber \\
    &\qquad \times \sum_{k=1}^i \Bigg[\bva H_R :\phi^{i-1}\left(\left\{\vec{x}\right\}\right): \phi^r\left(\left\{\vec{z}\right\}\right) \kva \left[i \delta^3\left(\abs{\vec{x}_k - \vec{z}}\right)\right] + \mathcal{O}\left(\gamma^2\right) \, .
\end{align}
With our knowledge of the functions $f_n\left(\left\{\vec{x}_n\right\}\right)$ with one of the arguments to be fixed at $\vec{z}$, we may explicitly integrate them against the above kernels to find the quantity we defined above to be $Z$ at any desired order of $\gamma$. In this paper, to avoid this subtlety, we will simply assume that the states have no support near the boundary of the ball. So, we will blithely ignore such terms. Therefore, yet again, we are only left with the second term with the Heaviside-Theta function in \eqref{eqn:boundary-correlators2}:
\begin{align}\label{eqn:boundary-correlators3}
    &\bpsi H_R \pi\left(\vec{z}\right) \phi^r\left(\left\{\vec{z}\right\}\right) \kpsi - \bva H_R \pi\left(\vec{z}\right) \phi^r\left(\left\{\vec{z}\right\}\right) \kva ~ = ~ \nonumber \\
    &\quad = ~ - i \gamma \int \sum_{i = 1}^\infty \left(\prod_{j=1}^{i} \mathrm{d}^3 \vec{x}_j\right) f_i\left(\left\{\vec{x}\right\}\right) \nonumber \\
    &\hspace{0.75cm} \times i \sum_{k=1}^i \Bigg[ \bva :\phi^{i-1}\left(\left\{\vec{x}\right\}\right): \pi\left(\vec{x}_k\right) \pi\left(\vec{z}\right) \phi^r\left(\left\{\vec{z}\right\}\right) \kva \Theta\left( R - \abs{\vec{x}_k}\right) \Bigg] + \mathcal{O}\left(\gamma^2\right) \, .
\end{align}
In this equation, all the $\mathcal{O}\left(\gamma^2\right)$ terms are now obtained by commuting the field insertions arising from the state with the r.h.s of this equation. This results in an expression with a single remnant conjugate momentum term, since the other one would result in a Dirac-delta after commutation: 
\begin{align}\label{eqn:boundary-correlators4}
    &\mathcal{O}\left(\gamma^2\right) ~ = ~ \left(- i \gamma\right)^2 \int \sum_{i = 1}^\infty \left(\prod_{j=1}^{i} \mathrm{d}^3 \vec{x}_j\right) f_i\left(\left\{\vec{x}\right\}\right) \int \sum_{p = 1}^\infty \left(\prod_{q=1}^{p} \mathrm{d}^3 \vec{x}_q\right) f_p\left(\left\{\vec{x}\right\}\right) \nonumber \\
    &\hspace{2cm} \times i^2 \sum_{s=1}^p \sum_{k=1}^i \Bigg[ \bva :\phi^{p-1}\left(\left\{\vec{x}\right\}\right): :\phi^{i-1}\left(\left\{\vec{x}\right\}\right): \Big[ \delta^3\left(\vec{x}_s - \vec{x}_k\right) \pi\left(\vec{z}\right) \nonumber \\
    &\hspace{5cm} + \delta^3\left(\vec{x}_s - \vec{z}\right) \pi\left(\vec{x}_k\right) \Big] \phi^r\left(\left\{\vec{z}\right\}\right) \kva \Theta\left( R - \abs{\vec{x}_k}\right) \Bigg] \, .
\end{align}
Since contractions of conjugate momenta with the field are identically vanishing on the $t=0$ slice (see \eqref{eqn:massivephipi}), we see all terms of $\mathcal{O}\left(\gamma^2\right)$ are all identically vanishing. One may be worried about the $\mathcal{O}\left(\gamma^3\right)$ terms which will now turn the conjugate momenta in the above expression into Dirac-delta functions leaving no more conjugate momenta in the matrix elements. All such terms would therefore necessarily have contact terms of the kind $\delta^3\left(\vec{x}_i - \vec{z}\right)$. As described in \secref{sec:states}, we will assume that there exists a radial location $R$ where the state has no local support, allowing us to drop all such contact terms. This also ensures that all terms in $Z$ also drop out since they come with similar contact terms as can be seen from \eqref{eqn:Z}. Therefore, yet again, we have that the linear answer is exact:
\begin{align}\label{eqn:boundary-correlators}
    &\bpsi H_R \pi\left(\vec{z}\right) \phi^r\left(\left\{\vec{z}\right\}\right) \kpsi - \bva H_R \pi\left(\vec{z}\right) \phi^r\left(\left\{\vec{z}\right\}\right) \kva ~ = ~ \nonumber \\
    &\hspace{1.5cm} = ~ - i \gamma \int \sum_{i = 1}^\infty \left(\prod_{j=1}^{i} \mathrm{d}^3 \vec{x}_j\right) f_i\left(\left\{\vec{x}\right\}\right) \nonumber \\
    &\hspace{2.5cm} \times i \sum_{k=1}^i \Bigg[ \bva :\phi^{i-1}\left(\left\{\vec{x}\right\}\right): \pi\left(\vec{x}_k\right) \pi\left(\vec{z}\right) \phi^r\left(\left\{\vec{z}\right\}\right) \kva \Theta\left( R - \abs{\vec{x}_k}\right) \Bigg] \, .
\end{align}
Finally, since all contractions of the conjugate momentum with the field itself are vanishing on an equal time slice, the fields must contract with each other rendering this correlator only non-vanishing when $r = i-1$, as we found in the bulk correlators too.

\subsection{Determining the one-particle contribution}\label{sec:f1}
From the general result \eqref{eqn:boundary-correlators}, the one-point boundary correlator is given by choosing $r=0=i-1$ to find
\begin{align}\label{eqn:f1}
    \bpsi H_R \pi\left(\vec{z}\right) \kpsi - \bva H_R \pi\left(\vec{z}\right) \kva ~ &= ~ \gamma \int \mathrm{d}^3 \vec{x} f_1\left(\vec{x}\right) \bva \pi\left(\vec{x}\right) \pi\left(\vec{z}\right) \kva \Theta\left( R - \abs{\vec{x}}\right) \nonumber \\
    &= \dfrac{- m^2 \gamma}{4 \pi^2} \int \mathrm{d}^3 \vec{x} f_1\left(\vec{x}\right) \dfrac{K_2\left(m \left|\vec{x} - \vec{z}\right|\right)}{\left|\vec{x} - \vec{z}\right|^2} \Theta\left( R - \abs{\vec{x}}\right) \, ,
\end{align}
where we used \eqref{eqn:massivepipi}. It is evident that the $\Theta$-function ensures that any inversion of this equation will only yield the information about the state inside the ball of radius $R$. In \appref{app:boundaryinversion}, we demonstrate how this integral can be inverted for the function $f_1\left(\vec{x}\right)$ yielding exactly the same information as we find from the bulk correlator \eqref{eqn:one-particle-info}.

\subsection{Determining the two-particle contribution}\label{sec:f2}
In analogy with the previous calculation, let us now consider the following correlator $ \langle \psi | H_{\text{sm}} \pi \left(z\right) \phi \left(z_{1}\right) | \psi \rangle$ using \eqref{eqn:boundary-correlators} with $r = i - 1 = 1$:
\begin{align}
    &\bpsi H_R \pi\left(\vec{z}\right) \phi\left(\vec{z}_1\right) \kpsi - \bva H_R \pi\left(\vec{z}\right) \phi\left(\left\{\vec{z}_1\right\}\right) \kva ~ = ~ \nonumber \\
    &\hspace{1cm} = ~ \gamma \int \mathrm{d}^3 \vec{x}_1 \mathrm{d}^3 \vec{x}_2 f_2\left(\vec{x}_1 , \vec{x}_2\right) \Bigg[ \bva \phi\left(\vec{x}_1\right) \pi\left(\vec{x}_2\right) \pi\left(\vec{z}\right) \phi\left(\vec{z}_1\right) \kva \Theta\left( R - \abs{\vec{x}_1}\right) + \nonumber \\
    &\hspace{5.5cm} + \bva \phi\left(\vec{x}_2\right) \pi\left(\vec{x}_1\right) \pi\left(\vec{z}\right) \phi\left(\vec{z}_1\right) \kva \Theta\left( R - \abs{\vec{x}_2}\right) \Bigg] \nonumber \\
    &\hspace{1cm} = ~ \gamma \int \mathrm{d}^3 \vec{x}_1 \mathrm{d}^3 \vec{x}_2 \tilde{f}_2\left(\vec{x}_1 , \vec{x}_2\right) \Bigg[ \bva \phi\left(\vec{x}_1\right) \pi\left(\vec{x}_2\right) \pi\left(\vec{z}\right) \phi\left(\vec{z}_1\right) \kva \Theta\left( R - \abs{\vec{x}_1}\right) \Bigg] \, ,
\end{align}
where in the last step, we defined $\tilde{f}_2\left(\vec{x}_1 , \vec{x}_2\right) \coloneqq f_2\left(\vec{x}_1 , \vec{x}_2\right) + f_2\left(\vec{x}_2 , \vec{x}_1\right)$. Since contractions of the conjugate momentum with the field are vanishing in the vacuum, the only allowed Wick-contraction is between the two conjugate momenta and between the two field insertions. This results in 
\begin{align}
    &\bpsi H_R \pi\left(\vec{z}\right) \phi\left(\vec{z}_1\right) \kpsi - \bva H_R \pi\left(\vec{z}\right) \phi\left(\left\{\vec{z}_m\right\}\right) \kva ~ = ~ \nonumber \\
    & ~ = ~ \dfrac{- m^3 \gamma}{16 \pi^4} \int \mathrm{d}^3 \vec{x}_1 \mathrm{d}^3 \vec{x}_2 \tilde{f}_2\left(\vec{x}_1 , \vec{x}_2\right) \Bigg[ \dfrac{K_2\left(m \left|\vec{x}_2 - \vec{z}\right|\right)}{\left|\vec{x}_2 - \vec{z}\right|^2} \dfrac{K_1\left(m \left|\vec{x}_1 - \vec{z}_1\right|\right)}{\left|\vec{x}_1 - \vec{z}_1\right|} \Theta\left( R - \abs{\vec{x}_1}\right) \Bigg] \, .
\end{align}
As we argue in \appref{app:boundaryinversion}, this integral equation can again be inverted to find precisely the same information about the two-particle contribution as we did int he bulk correlator \eqref{eqn:bulktwoptfn}. Namely, we find that this integral can be inverted to find the function $\tilde{f}_2\left(\vec{x}_1 , \vec{x}_2\right)$ but with $\vec{x}_1$ restricted to be inside the sphere. Since the function is symmetrised, this implies that we find the complete function if one of the arguments is restricted to be inside the sphere.

\subsection{Determining all multi-particle contributions}
Having determined $f_{1}$ in \secref{sec:f1} and $f_{2}$ in \secref{sec:f2}, we now move on to contributions from multi-particle exponents in \eqref{eqn:basisState}. Indeed, it is straightforward to see that a correlation function with three insertions such as $\langle \psi | H_{\text{sm}} \pi \left(\vec{z}\right) \phi \left(\vec{z}_{1}\right) \phi \left(\vec{z}_{2}\right) | \psi \rangle$ generically contains contributions from both $f_{1}$ and $f_{3}$ and captures the bulk correlator arising from $\bpsi \pi\left(\vec{y}_1\right) \phi \left(\vec{x}_{2}\right) \phi \left(\vec{x}_{3}\right) \kpsi$. Having already determined $f_{1}$, this correlator allows us to identify $f_{3}$. Similarly, a correlator of the form $\bpsi | H_{\text{sm}} \pi \left(\vec{z}\right) \phi \left(\vec{z}_{1}\right) \phi \left(\vec{z}_{2}\right) \phi \left(\vec{z}_{3}\right) \kpsi$ together with the knowledge of $f_{2}$ allows us to determine $f_{4}$ resulting in the information contained in the correlator $\bpsi | \pi \left(\vec{y}_{1}\right) \phi \left(\vec{x}_{2}\right) \phi \left(\vec{x}_{3}\right) \phi \left(\vec{x}_{4}\right) \kpsi$. Proceeding in this manner, we may algorithmically determine all multi-particle contributions, thereby identifying the state \eqref{eqn:basisState} to all orders in $\gamma$.

\section{Most general unitary excitations on the vacuum}\label{sec:generalStates}

In all of the calculations so far, we have restricted ourselves to a sub-space of states of the form \eqref{eqn:basisState}. While this is an infinite class of states parameterised by an infinite set of functions $f_n$, we know that the most general states contain insertions of the conjugate momentum too and the class is (at least morally speaking) doubly infinite, parameterised by functions $f_{n, m}$ as shown in \eqref{eqn:generalbasisstates}. Owing to the extra insertions of the conjugate momentum in the states, none of the correlation functions terminate at any finite order in $\gamma$. Therefore, even identifying what information the bulk correlators contain about the functions $f_{n,m}$, that define the state, is difficult to quantify. 

However, in this section, we show that at linear order in $\gamma$, the bulk correlators can indeed be reproduced by certain boundary correlators, lending credibility to the generality of the principle. In order to do so, we will consider an $\mathcal{O}\left(\gamma\right)$ expansion of the general states up to two particles (this can be done to incorporate any finite $n$-particle scenario, but with more tedious work)
\begin{align}\label{eqn:generalTwoParticleState}
    \kpsi ~ = ~ \exp \left( i \gamma \int \left(f_{1,0} \phi + f_{0,1} \pi + f_{2,0} : \phi^ 2 : + f_{1,1} : \phi \pi : + f_{0,2} : \pi^2 : \right)\right) \kva \, ,
\end{align}
where the arguments of the functions and fields are left implicit. Furthermore, to avoid clutter of notation, we will not write the subtraction of the vacuum correlators in what follows (this is left implicit).

\subsection*{One-point functions}
The bulk one-point functions in this state can easily be computed to give
\begin{align}
    \bpsi \pi\left(\vec{y}\right) \kpsi ~ &= ~ \gamma f_{1,0}\left(\vec{y}\right) + \mathcal{O}\left(\gamma^2\right) \quad \text{and} \\
    \bpsi \phi\left(\vec{y}\right) \kpsi ~ &= ~ - \gamma f_{0,1}\left(\vec{y}\right) + \mathcal{O}\left(\gamma^2\right) \, .
\end{align}
The corresponding boundary one-point functions are then
\begin{align}
    \bpsi H_R \pi\left(\vec{z}\right) \kpsi ~ &= ~ \gamma \int f_{1,0}\left(\vec{x}\right) \wick{\c\pi\left(\vec{x}\right) \c\pi\left(\vec{z}\right)} \Theta\left(R - \abs{\vec{x}}\right) \quad \text{and} \\
    \bpsi H_R \phi\left(\vec{z}\right) \kpsi ~ &= ~ \gamma \int  f_{0,1}\left(\vec{x}\right) \wick{\c\partial_t \pi \left(\vec{x}\right) \c\phi\left(\vec{z}\right)} \Theta\left(R - \abs{\vec{x}}\right) \, .
\end{align}
Here, the last Wick-contraction is a time derivative of the expression we found in \eqref{eqn:massivephipi}:
\begin{equation}
    \wick{\c\partial_t \pi \left(\vec{x}\right) \c\phi\left(\vec{y}\right)} ~ = ~ \dfrac{m^2}{4 \pi^2 \abs{\vec{x} - \vec{y}}^2} K_2\left(m \abs{\vec{x} -\vec{y}}\right) \, .
\end{equation}
Applying the inversion algorithm in \appref{app:boundaryinversion} that we have already used multiple times, we see that these boundary correlators allow us to find exactly the same information as the bulk correlators above since there, $\vec{y}$ is restricted to be inside the ball by definition of a bulk measurement.

\subsection*{Two-point functions}
There are three inequivalent bulk two-point functions at hand, all of which are non-trivial in the general state \eqref{eqn:generalTwoParticleState}. These are given by
\begin{align}
    \bpsi \phi\left(\vec{y}_1\right) \phi\left(\vec{y}_2\right) \kpsi ~ &= ~ -\gamma \int f_{1,1}\left(\vec{x}_1 , \vec{y}_1\right) \wick{\c1\phi\left(\vec{x}_1\right) \c1\phi\left(\vec{y}_2\right)} + \left(\vec{y}_1 \leftrightarrow \vec{y}_2\right) + \mathcal{O}\left(\gamma^2\right) \\ 
    \bpsi \pi\left(\vec{y}_1\right) \pi\left(\vec{y}_2\right) \kpsi ~ &= ~ \gamma \int f_{1,1}\left(\vec{y}_1 , \vec{x}_2\right) \wick{\c1\pi\left(\vec{x}_2\right) \c1\pi\left(\vec{y}_2\right)} + \left(\vec{y}_1 \leftrightarrow \vec{y}_2\right) + \mathcal{O}\left(\gamma^2\right) \\ 
    \bpsi \pi\left(\vec{y}_1\right) \phi\left(\vec{y}_2\right) \kpsi ~ &= ~ \gamma \int \tilde{f}_{2,0}\left(\vec{y}_1 , \vec{x}_2\right) \wick{\c1\phi\left(\vec{x}_2\right) \c1\phi\left(\vec{y}_2\right)} - \gamma \int \tilde{f}_{0,2}\left(\vec{y}_2 , \vec{x}_2\right) \wick{\c1\pi\left(\vec{x}_2\right) \c1\pi\left(\vec{y}_1\right)} \label{eqn:general_piphibulk} \\
    &\qquad \qquad + \mathcal{O}\left(\gamma^2\right) \, , \nonumber
\end{align}
where the tilde on the functions refers to a symmetrisation in $\vec{x}_1$ and $\vec{x}_2$. Of course, a fourth correlator $\bpsi \phi\left(\vec{z}_1\right) \pi\left(\vec{z}_2\right) \kpsi$ can be obtained from the third one above. The corresponding boundary two-point functions are then given, respectively, by the standard prescription with an insertion of the finite-radius Hamiltonian
\begin{align}
    \bpsi H_R \phi\left(\vec{z}_1\right) \phi\left(\vec{z}_2\right) \kpsi ~ &= ~ -\gamma \int f_{1,1}\left(\vec{x}_1 , \vec{x}_2\right) \wick{\c1\phi\left(\vec{x}_1\right) \c1\phi\left(\vec{z}_1\right)} \wick{\c1\partial_t \pi\left(\vec{x}_2\right) \c1\phi\left(\vec{z}_2\right)} \Theta\left(R - \abs{\vec{x}_2}\right) \nonumber \\
    &\qquad\qquad + \left(\vec{z}_1 \leftrightarrow \vec{z}_2\right) + \mathcal{O}\left(\gamma^2\right) \\ 
    \bpsi H_R \pi\left(\vec{z}_1\right) \pi\left(\vec{z}_2\right) \kpsi ~ &= ~ \gamma \int f_{1,1}\left(\vec{x}_1 , \vec{x}_2\right) \wick{\c1\pi\left(\vec{x}_1\right) \c1\pi\left(\vec{z}_1\right)} \wick{\c2\pi\left(\vec{x}_2\right) \c2\pi\left(\vec{z}_2\right)} \Theta\left(R - \abs{\vec{x}_1}\right) \\
    &\qquad\qquad + \left(\vec{z}_1 \leftrightarrow \vec{z}_2\right) + \mathcal{O}\left(\gamma^2\right) \nonumber \\ 
    \bpsi H_R \pi\left(\vec{z}_1\right) \phi\left(\vec{z}_2\right) \kpsi ~ &= ~ \gamma \int \tilde{f}_{2,0}\left(\vec{x}_1 , \vec{x}_2\right) \wick{\c1\pi\left(\vec{x}_1\right) \c1\pi\left(\vec{z}_1\right)} \wick{\c2\phi\left(\vec{x}_2\right) \c2\phi\left(\vec{z}_2\right)} \Theta\left(R - \abs{\vec{x}_1}\right) \nonumber \\
    &~ - \gamma \int \tilde{f}_{0,2}\left(\vec{x}_1 , \vec{x}_2\right) \wick{\c1\partial_t \pi\left(\vec{x}_1\right) \c1\phi\left(\vec{z}_2\right)} \wick{\c2\pi\left(\vec{x}_2\right) \c2\pi\left(\vec{z}_1\right)} \Theta\left(R - \abs{\vec{x}_1}\right) \label{eqn:general_piphiboundary}\\
    &\qquad \qquad + \mathcal{O}\left(\gamma^2\right) \, . \nonumber
\end{align}
It is no longer clear what information about the functions outside the ball a moment expansion allows us to invert for. But, expanding the functions in a basis as explained towards the end of \appref{app:bulkinversion}, we may obtain a set of linear equations to solve the system. However, even without having to invert, it is easy to see that both bulk and boundary correlators contain the same information about the state. As an example, let us consider \eqref{eqn:general_piphibulk} and \eqref{eqn:general_piphiboundary}. First, it is evident that the kernels conjugating the functions $f_{2,0}$ and $f_{0,2}$ in both the bulk and boundary integrals are identical for the argument $x_2$. These are the field and conjugate momentum two-point functions respectively. Moreover, the two functions appear in the same linear combination in both equations. Therefore, both the bulk and boundary equations yield the same information about the functions in terms of one variable. The bulk equation gives us a linear combination of the functions in terms of the second variable but restricted to be inside the ball. Whereas from the boundary equation, owing to the step function, we may extract exactly the same information.

Therefore, even without having to invert these expressions for the functions defining the state, it is nevertheless evident that the boundary correlators allow for exactly the same information about the state to be extracted.

\section{Discussion and shortcomings}\label{sec:discussion}

In this article, we considered linearised gravity coupled minimally to a free Klein-Gordon scalar field theory. We defined a certain basis of states that are excitations built out of unitary operators acting on the vacuum in flat space. These states were motivated by an interest in understanding a notion of holography of information in finite regions of space. We chose a ball of radius $R$ to this end. We showed that all information that can be extracted from bulk correlators inside the ball can also be extracted from the boundary of the ball\footnote{provided that we allow ourselves measurements in a small radial interval near the boundary}, to linear order in $\gamma$ for the states of the form $\kpsi = U \kva$, created by the most general unitary operators available in the theory as defined in \eqref{eqn:generalUnitary}. We showed that bulk correlators determine the dependence on the functions $f_n$ (that define the said states) on all but one argument everywhere on the Cauchy slice, while its dependence on one argument is restricted to be within the ball. We argued that the knowledge of the functions outside the ball is owed to non-trivial entanglement between the inside and outside of the ball. We explicitly showed that the same information about the functions can be determined from boundary correlators too. We then defined an interesting infinite sub-class of states in \eqref{eqn:basisState}. For this infinite sub-class of states, our results are valid non-perturbatively in the parameter $\gamma$ appearing in the states. These results lend credibility to a notion of holography of information in finite regions of spacetime, without having to go to infinity, at least within the confines of our setup.

However, there are several obvious shortcomings of our work that warrant mention. The primary shortcoming of our presentation is that our results are valid to linear order in $\gamma$ for the most general unitaries discussed in \secref{sec:generalStates}. Our technology fails to give us conclusive answers either way at higher orders. It would be a significant improvement to find a method to determine whether the principle of holography of information in a ball of finite radius holds to all orders in perturbation theory. We discuss several other issues of concern and possible improvements in what follows. 

\subsection{Non-linearities and field-theory interactions}

We have restricted our attention to linearised gravity in this article. Therefore, all states we considered are only built out of scalar operators. However, it is rather straightforward (even if slightly more cumbersome) to include free gravitons in the theory. This requires us to include the gravitational stress-tensor in \eqref{eqn:HRasenergy}, which we ignored. Since our construction for finite region boundary observables that contain the bulk information involves a special choice of basis of states and corresponding correlators, this may easily be extended to the case of graviton field insertions and its corresponding conjugate momentum insertions. However, going beyond quadratic gravitational interactions is a significant challenge. It comes with further challenges that entail the positivity (or potential lack thereof) of the finite-radius Hamiltonian. 

Although we have not incorporated these effects in the present article, a self-interacting Klein-Gordon field should pose no conceptual problems. The kernels that are involved in inverting for the information of the states from the (bulk/boundary) correlators will now be significantly more involved, depending on the nature of the interactions. Nevertheless, we expect the conceptual conclusions to remain unchanged.

\subsection{Gauge invariance and gravitational dressing}

In the presentation so far, we have fixed a finite radius in the bulk to define our local region of interest. This is of course an ill-defined quantity because it is entirely gauge dependent. Of course, we expect that physical quantities measured in any gauge should all be consistent with each other. Strictly speaking, there are no local gauge invariant observables in quantum gravity. At the leading order in interactions that we concerned ourselves with, this is not a concern. However, when non-linearities discussed above are taken into consideration, observables can be defined in a gauge invariant manner by defining them relationally to points on the boundary of spacetime. These observables are gravitationally dressed in that they are no strictly localised owing to non-vanishing Poisson brackets (commutators in the quantum theory) at spacelike separation. This notion goes back to Brown and Kuchar \cite{Brown:1994py} and has recently been discussed in de Sitter space \cite{Chakraborty:2023los, Chakraborty:2023yed, Kaplan:2024xyk}.

In the present context, let us pick a point, say $u_0$ which is finite, on $\mathcal{I}^+$ (future null-infinity). Let us consider a Cauchy slice defined by the vanishing of the affine parameter associated with the null-rays of the massless matter field $\phi$. This matter can be taken to be some heavy dust shell. We now trace null rays of the massless matter field $\phi$ emanating from the boundary point $u$ back into the bulk. We define the finite region of interest to be that which is picked out by the intersection of this ray with the said Cauchy slice. This gives a relational definition of a finite region in the bulk that is gauge invariant. With this relational definition of a finite region in the bulk, we may have liked to formulate the principle of holography of information of a finite region as the statement that all information of localised (perturbative) states inside a ball of finite size in asymptotically flat space is also available on the boundary of the said ball. And that the same information can be represented in a gauge invariant manner on $\mathcal{I}^+$ as all information that is available at $u_0$.

However, the primary challenge in formulating such a principle at null-infinity in a gauge invariant manner is that a suitable analog of the smeared Hamiltonian at finite null time on $\mathcal{I}^+$ is unknown. One may have considered the Bondi mass aspect, but that operator is known to have infinite quadratic fluctuations \cite{Bousso:2017xyo, Witten:2021unn, Ciambelli:2024swv}. 

An alternative may be to consider dressing the operators in the bulk directly (which compromises on their locality) in the spirit of \cite{Donnelly:2015hta, Donnelly:2016rvo}, which we leave for future work.

\subsection{Unitary excitations on general Hermitean states}

The unitary excitations of the present article naturally have unit norm. The most general excitations on top of the vacuum, as written in \eqref{eqn:generalbasisstates} contain Hermitean pieces such that 
\begin{align}\label{eqn:basisState_hermitean}
    |\psi\rangle ~ &\coloneqq ~ e^{-X} \kg ~ = ~ \exp\left( i \gamma \sum_{n=0}^\infty \int \left( \prod_{i=1}^n\mathrm{d}^3 \vec{x}_i \right) \, f_n \left(\left\{\vec{x}\right\}\right) : \phi^n \left(\left\{x\right\}\right) : \right) \kg \, ,
\end{align}
where we defined
\begin{align}
    \kg ~ &\coloneqq ~ \exp\left( \gamma \sum_{n=0}^\infty \int \left( \prod_{i=1}^n \mathrm{d}^3 \vec{x}_i \right) \, g_n \left(\left\{\vec{x}\right\}\right) : \phi^n \left(\left\{x\right\}\right) : \right) \kva \, .
\end{align}
Here, the functions $g_n \left(\left\{\vec{x}\right\}\right)$ can also taken to be real because any complex parts can be absorbed into the functions $f_n$. It is then straightforward to repeat the exercise leading up to \eqref{eqn:bulk-correlators} in these general states to find
\begin{align}\label{eqn:bulk-correlators-hermitean}
    &\bpsi \pi\left(\vec{y}\right) \phi^m\left(\left\{\vec{y}\right\}\right) \kpsi - \bva \pi\left(\vec{y}\right) \phi^m\left(\left\{\vec{y}\right\}\right) \kva ~ = ~ \nonumber \\
    &\quad = - i \gamma \int \sum_{i = 1}^\infty \left(\prod_{j=1}^{i} \mathrm{d}^3 \vec{x}_j\right) f_i\left(\left\{\vec{x}\right\}\right) \sum_{k=1}^i \bg :\phi^{i-1}\left(\left\{\vec{x}\right\}\right): \phi^m\left(\left\{\vec{y}\right\}\right) \kg \left[i \delta^3\left(\abs{\vec{x}_k - \vec{y}}\right)\right] \nonumber \\
    &\hspace{7cm} \text{where } \phi^{i-1}\left(\left\{\vec{x}\right\}\right) \text{has no insertion at } \vec{x}_k \, .
\end{align}
In contrast to the unitary case, this correlator has non-vanishing contributions from all values of $m$ and $i$ because the normal ordered insertions may contract with field operators arising from the state $\kg$. Therefore, these correlators do not isolate a specific multi-particle exponents and receive contributions from all multi-particle insertions in general. Worse still, while the correlators in the unitary case had known kernels which were vacuum correlators, we now have kernels which are correlators in the state $\kg$. 

\subsection{Excited states and black holes}

We have restricted our attention to background field perturbation theory about flat space in this article. As discussed in the previous subsection, perturbations about general Hermitean states are of general interest. In particular, an interesting case is that of black holes. It is expected that simple observations outside the black hole horizon are insufficient to decode information from the black hole interior. Several observables may be calculated in the near-horizon region \cite{Gaddam:2020rxb, Gaddam:2020mwe, Gaddam:2021zka, Gaddam:2022pnb, Feleppa:2023eoi}. Moreover, black hole states may be represented by typical states as described in \cite{Marolf:2015dia} for example. In either of these setups, it would be very interesting to find at least a special class of unitary excitations on top of the black hole background that may be decoded by simple measurements near the horizon. It must however be emphasised that we defined states on a fixed time slice in this article. In this sense, this problem may not be unique to black holes and may be generic to generic states that are sufficiently complicated as discussed in the previous sub-section. In this article, we have not addressed these subtleties and leave them for future work.

\section*{Acknowledgements}
We are grateful to Chandramouli Chowdhury for collaboration in the initial stages of this work. It is a pleasure to thank Alok Laddha, R Loganayagam, Suvrat Raju, Ashoke Sen, and Omkar Shetye for various helpful discussions throughout the course of this work. We are also grateful to the string theory and quantum gravity group in ICTS. We acknowledge the support of the Department of Atomic Energy, Government of India, under project no. RTI4001.

\appendix

\section{Quadratic fluctuations of the Hamiltonian}\label{app:Hsmfluctuations}
Consider the mode expansion of a massive scalar field in asymptotically flat space in 3+1 dimensions with mostly plus signature
\begin{equation}
    \phi(x) ~ = ~ \int \dfrac{\mathrm{d}^3 k}{\sqrt{\left(2\pi\right)^3 2 \omega_{\vec{k}}}} \left(a_{\vec{k}} e^{i k\cdot x} + a_{\vec{k}}^{\dagger} e^{-i k\cdot x} \right) \quad \text{with} \quad \left[a_{\vec{k}} , a^{\dagger}_{\vec{k}'}\right] ~ = ~ \delta^{(3)}\left(\vec{k} - \vec{k}'\right) \, .
\end{equation}
The corresponding Hamiltonian density is then given by
\begin{align}
    \mathcal{H}\left(t,\vec{x}\right) ~ &\coloneqq ~ \dfrac{1}{2} : \left(\dot{\phi}^2 + \left(\nabla \phi\right)^2 \right) : \nonumber \\
    &= ~ - \dfrac{1}{2} \int \dfrac{\mathrm{d}^3 k \, \mathrm{d}^3 q  \left(\omega_{\vec{k}} \omega_{\vec{q}} + \vec{k} \cdot \vec{q} \right)}{2 \left(2\pi\right)^3 \sqrt{\omega_{\vec{k}} \omega_{\vec{q}}}} \Bigg[ a_{\vec{k}} a_{\vec{q}} e^{- i \left(\omega_{\vec{k}} + \omega_{\vec{q}}\right) t + i \left(\vec{k} + \vec{q}\right) \cdot \vec{x}} \nonumber \\
    &\qquad \qquad \qquad \qquad \qquad + a_{\vec{k}}^{\dagger} a^\dagger_{\vec{q}} e^{i \left(\omega_{\vec{k}} + \omega_{\vec{q}}\right) t - i \left(\vec{k} + \vec{q}\right) \cdot \vec{x}} - a_{\vec{k}}^{\dagger} a_{\vec{q}} e^{- i \left(\omega_{\vec{k}} + \omega_{\vec{q}}\right) t + i \left(\vec{k} - \vec{q}\right) \cdot \vec{x}} \nonumber \\
    &\qquad \qquad \qquad \qquad \qquad - a^\dagger_{\vec{q}} a_{\vec{k}} e^{i \left(\omega_{\vec{k}} + \omega_{\vec{q}}\right) t - i \left(\vec{k} - \vec{q}\right) \cdot \vec{x}} \Bigg] \, ,
\end{align}
Integrating this density over in a finite region of space from, say, $r = 0$ to $r= R$ allows us to define
\begin{align}\label{eqn:HRapp}
    H \left(R, t\right) ~ &\coloneqq ~ \int_{\left|\vec{x}\right| \leq R} \mathrm{d}^3 x \, \mathcal{H}\left(t, \vec{x}\right) ~ = ~ \int_{0}^R r^2 \mathrm{d}r \int_0^\pi \sin\left(\theta\right) \mathrm{d}\theta \int_0^{2 \pi} \mathrm{d} \phi \, \mathcal{H}\left(t, r, \theta, \phi\right) \nonumber \\
    &= ~ \dfrac{1}{2} \int \dfrac{\mathrm{d}^3 k \, \mathrm{d}^3 q  \left(\omega_{\vec{k}} \omega_{\vec{q}} + \vec{k} \cdot \vec{q} \right)}{\left(2\pi\right)^2 \sqrt{\omega_{\vec{k}} \omega_{\vec{q}}}} \Bigg[ a^\dagger_{\vec{k}} a_{\vec{q}} \mathcal{D}\left(k - q, R, t\right) - a_{\vec{k}} a_{\vec{q}} \mathcal{D}\left(k + q, R, t\right) + \text{h.c.} \Bigg] \, ,
\end{align}
where we defined 
\begin{align}
    \mathcal{D} \left(q, R, t\right) ~ \coloneqq ~ \mathcal{D} \left(q, R\right)  e^{- i q^0 t} ~ \coloneqq ~ \dfrac{ \sin\left(\left|\vec{q}\right| R\right) - \left|\vec{q}\right| R \cos\left(\left|\vec{q}\right| R\right) }{\left|\vec{q}\right|^3 } e^{- i q^0 t} \, .
\end{align}
We now smear the above operator \eqref{eqn:HRapp} in time in terms of a small length scale $\eta$, say about the $t=0$ slice, as
\begin{align}\label{eqn:Hsm}
    H_{\text{sm}} ~ \coloneqq ~ \int_{-\infty}^{+\infty} \mathrm{d}t \, \mathfrak{g}\left(t\right) H \left(R, t\right) \quad \text{with} \quad \mathfrak{g}\left(t\right) ~ \coloneqq ~  \dfrac{1}{\eta \sqrt{\pi}} \exp\left(-\dfrac{t^2}{\eta^2}\right) \, .
\end{align}
Carrying out the time-smeared integral in \eqref{eqn:Hsm}, we find
\begin{align}
    H_{\text{sm}} ~ &= ~ \dfrac{1}{2} \int \dfrac{\mathrm{d}^3 k \, \mathrm{d}^3 q  \left(\omega_{\vec{k}} \omega_{\vec{q}} + \vec{k} \cdot \vec{q} \right)}{\left(2\pi\right)^2 \sqrt{\omega_{\vec{k}} \omega_{\vec{q}}}} \Bigg[ a^\dagger_{\vec{k}} a_{\vec{q}} \tilde{\mathcal{D}}\left(k - q, R, \eta\right) - a_{\vec{k}} a_{\vec{q}} \tilde{\mathcal{D}}\left(k + q, R, \eta\right) + \text{h.c.} \Bigg] \, ,
\end{align}
where we now defined
\begin{align}
    \tilde{\mathcal{D}} \left(q, R, \eta\right) ~ \coloneqq ~ \mathcal{D} \left(q, R\right)  \exp\left(- \dfrac{\left(q^0\right)^2 \eta^2}{4}\right) \, .
\end{align}
Therefore, we see that the vacuum expectation value of the quadratic fluctuations of this smeared operator takes the form
\begin{align}
    \langle \Omega| H^2_{\text{sm}} |\Omega\rangle ~ = ~ \int \mathrm{d}t_1 \mathrm{d}t_2 \, \mathfrak{g}\left(t_1\right) \mathfrak{g}\left(t_2\right) H \left(R, t_1\right) H \left(R, t_2\right) \, .
\end{align}
Since
\begin{equation}
    \int_{-\infty}^{+\infty} \mathrm{d}t \, \mathfrak{g}\left(t\right) e^{\pm i b t} ~ = ~ \exp\left(- \dfrac{b^2 \eta^2}{4}\right) \, , 
\end{equation}
we see that the quadratic fluctuations of the momentum integrals in the smeared Hamiltonian at finite radius converge owing to the exponential suppression. 

In the large $R$ limit, however, the operator is still divergent. To deal with this, in \cite{Laddha:2020kvp}, an additional radial smearing was proposed to ensure that the $R \rightarrow \infty$ limit does not result in divergent quadratic fluctuations for  massless scalar fields. A refined smearing can be employed to deal with the massive case. In this note, we only concern ourselves with the determination of a localised state inside a ball of the fixed radius $R$ in terms of correlation functions on the boundary of the said ball, with no large $R$ limit. To this end, a time-smeared Hamiltonian suffices. 

\section{Finite-radius Hamiltonian as a boundary operator}\label{app:HRboundaryterm}

In this section, we review the standard result that the Hamiltonian defined in \eqref{eqn:HR} is indeed a boundary term on the surface of the ball. To this end, let us consider the Einstein's equations in the presence of matter sources:
\begin{equation}
R_{\mu\nu} - \dfrac{1}{2} g_{\mu\nu} R ~ = ~ 8 \pi G T_{\mu\nu} \, .
\end{equation}
Perturbing the metric about flat space as
\begin{equation}
g_{\mu \nu} ~ = ~ \eta_{\mu\nu} + h_{\mu\nu} \, , 
\end{equation}
we have the following first order variations of its derivatives
\begin{align}
\Gamma^{\mu(1)}_{\nu\lambda} ~ &= ~ \dfrac{1}{2} \eta^{\mu\nu} \left( \partial_{\nu} h_{\rho\lambda} + \partial_{\lambda} h_{\rho\nu} - \partial_{\lambda} h_{\nu\lambda}  \right) \nonumber \\
R^{(1)}_{\mu\nu} ~ &= ~ \dfrac{1}{2} \left(\partial^{\lambda} \partial_{\mu} h_{\lambda\nu} + \partial^{\lambda} \partial_{\nu} h_{\lambda\mu} - \partial^2 h_{\mu\nu} - \partial_{\mu} \partial_{\nu} h \right) \nonumber \\ 
R^{(1)} ~ &= ~ \partial^{\lambda} \partial^{\rho} h_{\lambda\rho} - \partial^2 h \, .
\end{align}
As is common practice, we defined the quantities $h \coloneqq \eta^{\mu\nu} h_{\mu\nu}$ and $\partial^2 \coloneqq \partial^{\mu} \partial_{\mu}$. Considering higher order variations, we may in general write
\begin{align}
R_{\mu\nu} ~ = ~ R^{(1)}_{\mu\nu} + \sum_n R^{(n)}_{\mu\nu} \qquad \text{and} \qquad R ~ = ~ R^{(1)} + \sum_n R^{(n)} \, .
\end{align}
The summation here includes all the non-linear terms in $h_{\mu\nu}$. In this notation, Einstein's equations can now be written as
\begin{align}\label{eqn:einsteineom}
R^{(1)}_{\mu\nu} - \dfrac{1}{2} \eta_{\mu\nu} R^{(1)} ~ &= ~ 8 \pi G T_{\mu\nu} +\dfrac{1}{2} h_{\mu\nu} R + \dfrac{1}{2} \eta_{\mu\nu} \sum_n R^{(n)} - \sum_n R^{(n)}_{\mu\nu} \nonumber \\
&\eqqcolon ~ 8 \pi G \left(T_{\mu\nu} + T^{h}_{\mu\nu} \right) \, ,
\end{align}
where we have defined all the non-linear terms in metric perturbations suggestively as as $T^{h}_{\mu\nu}$. The reason for doing so is the simple observation that
\begin{align}
\eta^{\mu\rho} \partial_{\rho} \left( R^{(1)}_{\mu\nu} - \dfrac{1}{2} \eta_{\mu\nu} R^{(1)} \right) ~ = ~ 0 \quad \text{which implies that} \quad \eta^{\mu\rho} \partial_{\rho} \left(T_{\mu\nu} + T^{h}_{\mu\nu} \right) ~ = ~ 0 \, .
\end{align}
Therefore, we may consider $T^{h}_{\mu\nu}$ as the energy-momentum tensor associated with the fluctuations $h_{\mu\nu}$. This allows us to now integrate both sides of the equations \eqref{eqn:einsteineom} inside the ball of radius $R$, using the unit vector in the radial direction $n_i$, as
\begin{align}
\int_{\abs{\vec{x}}\leq R} \mathrm{d}^3 \vec{x} \, \left(T_{00} + T^h_{00}\right) ~ &= ~ \dfrac{1}{8\pi G} \int_{\abs{\vec{x}}\leq R} \mathrm{d}^3 \vec{x} \left( R^{(1)}_{00} - \dfrac{1}{2} \eta_{00} R^{(1)} \right) \\
&= ~ \dfrac{1}{16\pi G} \int_{\abs{\vec{x}}\leq R} \mathrm{d}^3 \vec{x} \, \partial_i \left( \partial_{j} h_{ij} - \partial_i h_{jj} \right) \\
&= ~ \dfrac{1}{16\pi G} \int_{\abs{\vec{x}}=R} \mathrm{d}^2 x \, n_i \left( \partial_{j} h_{ij} - \partial_i h_{jj} \right) \, .
\end{align}

In this article, we focus on linearised gravity\footnote{It is an interesting exercise to extend our results to the non-linear gravitational theory which we expect to be straightforward at quadratic order.} and therefore ignore the energy-momentum tensor associated with the gravitons. With the knowledge that the energy-momentum tensor of a free Klein-Gordon scalar field is 
\begin{align}
    T_{\mu\nu} ~ = ~ \partial_\mu \phi \partial_\nu \phi - \dfrac{1}{2} g_{\mu\nu} \left(g^{\rho \sigma} \partial_\rho \phi \partial_\sigma \phi + m^2 \phi^2\right) \, ,
\end{align}
we see that 
\begin{align}\label{eqn:HRasenergy}
    \int_{\abs{\vec{x}} \leq R} \mathrm{d}^3 \vec{x} \, T_{00} ~ &= ~ \dfrac{1}{2} \int_{\abs{\vec{x}} \leq R} \mathrm{d}^3 \vec{x} \, \left(\dot{\phi}^2 + \left(\nabla\phi\right)^2 + m^2 \phi^2\right) \nonumber \\
    &= ~ \int_{\abs{\vec{x}} \leq R} \mathrm{d}^3 \vec{x} \, \mathcal{H} \left(t, \vec{x}\right) \nonumber \\
    &= ~ \dfrac{1}{16\pi G} \int_{\abs{\vec{x}}=R} \mathrm{d}^2 x \, n_i \left( \partial_{j} h_{ij} - \partial_i h_{jj} \right) \, . 
\end{align}
Therefore, we see that the finite-radius Hamiltonian defined in \eqref{eqn:HR} is indeed a boundary term in terms of metric fluctuations in linearised quantum gravity. Of course, we may include corrections to this by including non-linear terms in the metric fluctuations.

\section{Holography of information in a ball of infinite radius}

In this appendix, we will sketch how an observer at spatial infinity may construct the bulk state by measuring boundary correlation functions.\footnote{For a detailed discussion, we refer the reader to \cite{Chowdhury:2022wcv}.}

We start by noting that states defined on the bulk $t=0$ Cauchy slice are in a one to one correspondence with states defined on future null-infinity $\mathscr{I}^+$. Therefore, we may write a general state as,
\begin{equation}\label{eq:linear_basis_state}
\kpsi ~ = ~ \sum_{i=1}^{\infty} \int \left(\prod_{j=1}^i \mathrm{d}^3 \vec{x}_j\right) f_i(\left\{x_j\right\}) : \phi^{i} \left(\left\{x\right\}\right) : \kva \, .
\end{equation}
In this basis, we choose the field insertions to be normal ordered. These insertions are defined at $\mathscr{I}^+$ and can be obtained by taking the $r\rightarrow\infty$ limit of the bulk fields, keeping $t-r=u$ fixed.\footnote{More precisely, $\lim_{r \rightarrow \infty} \phi \left(r,t,\Omega\right)|_{\text{fixed } u} ~ = ~ \phi^{\text{bulk}} \left(u,\Omega\right) ~ = \frac{1}{r} \phi^{\text{bdry}} \left(u,\Omega\right)$.} A unique identification of the states requires us to determine all ${f_i}$. To this end, we may act with a unitary of the form
\begin{equation}
U_1 ~ = ~ 1 + i \int \mathrm{d}^3 \vec{x} g_1\left(\vec{x}\right) \phi\left(\vec{x}\right) + \mathcal{O} \left(g^2\right) ~ \sim ~ \int_{-\infty}^{-1/\epsilon} \mathrm{d} u \int \mathrm{d} \Omega' g_1\left(u,\Omega'\right) \phi\left(u,\Omega'\right) \, .
\end{equation}
In the above unitary, we neglect higher order terms $\mathcal{O}\left(g^2\right)$ and beyond. Given the state \eqref{eq:linear_basis_state}, we now define a obtain a new state $\tilde{\kpsi} = U_1 \kpsi$. We will now compute the expectation value of the vacuum projector $P_0=\kva\bva$ in this new state. Since our observations are being made on future null-infinity $\mathscr{I}^+$ in the region $u\in \left(- \infty, -\frac{1}{\epsilon} \right)$, we have access to the complete Hamiltonian and consequently the projector onto the vacuum. This gives,
\begin{align}
    \tilde{\bpsi} P_0 \tilde{\kpsi} ~ &= ~ \bva \left( \sum_{i=1}^{\infty} \int \left(\prod_{j=1}^i \mathrm{d} \vec{x}_j\right) f_i \left(\left\{\vec{x}_j\right\}\right) : \phi^{i} \left(\left\{\vec{x}_j\right\}\right) : \left(1 + i \int \mathrm{d}^3 \vec{y} g\left(\vec{y}\right) \phi\left(\vec{y}\right) \right)  \right) \kva \nonumber \\ 
    & \quad \times \bva \left( \left( 1 - i \int \mathrm{d}^3 \vec{y} g \left(\vec{y}\right) \phi\left(\vec{y}\right) \right) \sum_{i=1}^{\infty} \int \left(\prod_{j=1}^i \mathrm{d}^3 \vec{x}_j\right) f_i \left(\left\{\vec{x}_j\right\}\right) : \phi^{i} \left(\left\{\vec{x}_j\right\}\right) : \right) \kva \nonumber \\ 
    &= ~ \left( \int \mathrm{d}^3 \vec{x} \mathrm{d}^3 \vec{y} f_1\left(\vec{x}_1\right) g_1\left(\vec{y}\right) \bva \phi\left(\vec{x}_1\right) \phi\left(\vec{y}\right) \kva \right)^2 \, .
\end{align}
In this calculation, normal ordering in the choice of basis ensured that only the $f_1$ term in the state contributes to the correlator. Therefore, we had a straightforward choice of unitary ($U_1$) that isolates the $f_1$ contribution. Similarly, it is easy to verify that $U_2$ (which comes with a two-particle insertion) isolates the $f_2$ contribution while $U_3$ isolates $f_1$ and $f_3$. Therefore, starting with the lower-particle insertions, we may iteratively determine all multi-particle contributions to the state. From these correlators, we may use the algorithm in \appref{app:bulkinversion} to invert for the functions $f_i$ with the help of many measurements in the interval $u\in \left(- \infty, -\frac{1}{\epsilon} \right)$. As is evident, the role played by the vacuum projector is pivotal in easily isolating the individual contributions to the state.

\section{Bulk correlators}\label{app:bulkcorrelators}
Here, we compute the bulk correlators that define for us, information inside of a ball. Given some arbitrary spacetime points $x_1 = \left(x^0_1 , \vec{x}_1\right)$ and $x_2 = \left(x^0_2 , \vec{x}_2\right)$, the bulk scalar two-point correlator of interest is
\begin{align}
    \bva \phi\left(x_1\right) \phi\left(x_2\right) \kva ~ &= ~ \bva \phi\left(x^0_1 , \vec{x}_1\right) \phi\left(x^0_2 , \vec{x}_2\right) \kva \nonumber \\
    &= ~ \int \dfrac{\mathrm{d}^3 \vec{k} }{\sqrt{\left(2\pi\right)^3 2 \omega_{\vec{k}}}} \dfrac{\mathrm{d}^3 \vec{p} }{\sqrt{\left(2\pi\right)^3 2 \omega_{\vec{p}}}} e^{- i \omega_{\vec{k}} x^0_1} e^{i \omega_{\vec{p}} x^0_2} e^{i \vec{k} \cdot \vec{x}_1} e^{- i \vec{p} \cdot \vec{x}_2} \bva a_{\vec{k}} a^\dagger_{\vec{p}} \kva \nonumber \\
    &= ~ \int \dfrac{\mathrm{d}^3 \vec{k} }{\left(2\pi\right)^3 2 \omega_{\vec{k}}} e^{i \omega_{\vec{k}} \left(x^0_2 - x^0_1\right)} e^{i \vec{k} \cdot \left(\vec{x}_1 - \vec{x}_2\right)} \nonumber \\
    &= ~ \int \dfrac{k^2 \mathrm{d} k }{\left(2\pi\right)^2 \omega_{\vec{k}}} e^{i \omega_{\vec{k}} \left(x^0_2 - x^0_1\right)} \dfrac{\sin\left(k \left|\vec{x}_1 - \vec{x}_2\right|\right)}{k \left|\vec{x}_1 - \vec{x}_2\right|} \, , 
\end{align}
where $k \coloneqq \left|\vec{k}\right|$. To evaluate this integral, we consider the massless and massive cases separately.

\subsection{Massless fields}
In the case of massless scalar fields, we have that $\omega_{\vec{k}} = k$. Therefore, 

\begin{align}
    \bva \phi\left(x_1\right) \phi\left(x_2\right) \kva ~ &= ~ \int \dfrac{k^2 \mathrm{d} k }{\left(2\pi\right)^2 k} e^{i k \left(x^0_2 - x^0_1\right)} \dfrac{\sin\left(k \left|\vec{x}_1 - \vec{x}_2\right|\right)}{k \left|\vec{x}_1 - \vec{x}_2\right|} \nonumber \\
    &= ~ \dfrac{i}{2 \left(2\pi\right)^2 \left|\vec{x}_1 - \vec{x}_2\right|}\int \mathrm{d} k \left(e^{ i k \left(\left(x^0_2 - x^0_1\right) - \left|\vec{x}_1 - \vec{x}_2\right|\right)} - e^{ i k \left(\left(x^0_2 - x^0_1\right) + \left|\vec{x}_1 - \vec{x}_2\right|\right)}\right) \nonumber \\
    &= ~ \dfrac{-1}{8 \pi^2 \left|\vec{x}_1 - \vec{x}_2\right|} \left(\dfrac{1}{\left(x^0_2 - x^0_1\right) - \left|\vec{x}_1 - \vec{x}_2\right|- i \epsilon} - \dfrac{1}{\left(x^0_2 - x^0_1\right) + \left|\vec{x}_1 - \vec{x}_2\right| - i \epsilon}\right) \nonumber \\
    &= ~ \dfrac{ - 1}{4 \pi^2 \left[\left(x^0_2 - x^0_1 - i \epsilon\right)^2 - \left|\vec{x}_1 - \vec{x}_2\right|^2 \right]} \, , \label{eqn:phiphiBulk} 
\end{align}
where we introduced the usual $i \epsilon$ prescription. An alternative representation of this correlator can be obtained from the second last equality in the above equation to arrive at
\begin{align}
    \bva \phi\left(x_1\right) \phi\left(x_2\right) \kva ~ &= ~ \dfrac{-1}{8 \pi^2 \left|\vec{x}_1 - \vec{x}_2\right|} \mathcal{P}\left(\dfrac{1}{\left(x^0_2 - x^0_1\right) - \left|\vec{x}_1 - \vec{x}_2\right|}\right) \nonumber \\
    &\quad + \dfrac{1}{8 \pi^2 \left|\vec{x}_1 - \vec{x}_2\right|} \mathcal{P}\left(\dfrac{1}{\left(x^0_2 - x^0_1\right) + \left|\vec{x}_1 - \vec{x}_2\right|}\right) \nonumber \\ 
    &\quad - \dfrac{i \pi \delta\left(\left(x^0_2 - x^0_1\right) - \left|\vec{x}_1 - \vec{x}_2\right|\right)}{8 \pi^2 \left|\vec{x}_1 - \vec{x}_2\right|} + \dfrac{i \pi \delta\left(\left(x^0_2 - x^0_1\right) + \left|\vec{x}_1 - \vec{x}_2\right|\right)}{8 \pi^2 \left|\vec{x}_1 - \vec{x}_2\right|} \, .
\end{align}
This form allows us to write the commutator as
\begin{align}\label{eqn:masslessphiphicomm}
    \bva \left[\phi\left(x_1\right) , \phi\left(x_2\right)\right] \kva ~ &= ~ \dfrac{i \pi \delta\left(\left(x^0_2 - x^0_1\right) + \left|\vec{x}_1 - \vec{x}_2\right|\right)}{4 \pi^2 \left|\vec{x}_1 - \vec{x}_2\right|} - \dfrac{i \pi \delta\left(\left(x^0_2 - x^0_1\right) - \left|\vec{x}_1 - \vec{x}_2\right|\right)}{4 \pi^2 \left|\vec{x}_1 - \vec{x}_2\right|} \, .
\end{align}
Taking appropriate time-derivatives of \eqref{eqn:phiphiBulk}, we find
\begin{align}
    \bva \phi\left(x_1\right) \pi\left(x_2\right) \kva ~ &= ~ \dfrac{\left(x^0_2 - x^0_1\right)}{2 \pi^2 \left[\left(x^0_2 - x^0_1 - i \epsilon\right)^2 - \left|\vec{x}_1 - \vec{x}_2\right|^2 \right]^2} \label{eqn:masslessphipi}\\
    \bva \pi\left(x_1\right) \pi\left(x_2\right) \kva ~ &= ~ \dfrac{1}{2 \pi^2} \left( \dfrac{4 \left(x^0_2 -x^0_1\right)^2}{\left[(x^0_2 - x^0_1 - i \epsilon)^2 - \abs{\vec{x} - \vec{y}}^2\right]^3} - \dfrac{1}{\left[(x^0_2 - x^0_1 - i \epsilon)^2-\abs{\vec{x}-\vec{y}}^2\right]^2} \right) \, .
\end{align}

\subsection{Massive fields}\label{app:bulkcorrelators-massive}
The dispersion relation for a massive scalar field is $\omega^2_{\vec{k}} - k^2 = m^2$. It is convenient to parameterise the frequency and momentum as $k = m \sinh{\eta}$ and $\omega_{\vec{k}} = m \cosh{\eta}$. For ease of notation in the following integral, we also define $t\coloneqq x^0_2 - x^0_1$ and $r = \left|\vec{r}\right| = \left|\vec{x}_1 - \vec{x}_2\right|$. Consequently the two-point function is
\begin{align}
    \bva \phi\left(x_1\right) \phi\left(x_2\right) \kva ~ &= ~ \dfrac{1}{\left(2\pi\right)^2 r} \int_0^\infty \dfrac{k \mathrm{d} k }{\omega_{\vec{k}}} \Big(\cos \omega_{\vec{k}} t + i \sin \omega_{\vec{k}} t \Big) \sin\left(k r\right) \nonumber \\
    &= ~ \dfrac{-1}{\left(2\pi\right)^2 r} \dfrac{\partial}{\partial r} \int_0^\infty \mathrm{d} \eta \Big(\cos\left(m t \cosh\eta\right) + i \sin \left(m t \cosh \eta\right) \Big) \cos\left(m r \sinh{\eta}\right) \nonumber \\
    &\eqqcolon ~ \dfrac{-1}{8 \pi^2 r} \dfrac{\partial}{\partial r} \Big(\mathcal{I}_1 + i  \mathcal{I}_2\Big) \, ,
\end{align}
where we defined
\begin{align}
    \mathcal{I}_1 ~ &\coloneqq ~ 2 \int_0^\infty \mathrm{d} \eta \, \cos\left(m t \cosh\eta\right) \cos\left(m r \sinh{\eta}\right) \nonumber \\
    &= ~ \int_0^\infty \mathrm{d} \eta \, \Big(\cos\left(m t \cosh\eta + m r \sinh{\eta}\right) + \cos\left(m t \cosh\eta - m r \sinh{\eta}\right)\Big) \\
    \mathcal{I}_2 ~ &\coloneqq ~ 2 \int_0^\infty \mathrm{d} \eta \, \sin \left(m t \cosh \eta\right) \cos\left(m r \sinh{\eta}\right) \nonumber \\
    &= ~ \int_0^\infty \mathrm{d} \eta \, \Big(\sin\left(m t \cosh\eta + m r \sinh{\eta}\right) + \sin\left(m t \cosh\eta - m r \sinh{\eta}\right)\Big) \, .
\end{align}
At this point, we consider the cases of space-like and time-like insertions separately. 

\paragraph{Space-like separation: } In this case, we have that $\left(x_1 - x_2\right)^2 = \left(- t^2 + r^2\right) > 0$. Choosing a frame where $t=0$, the integrals $\mathcal{I}_{1}$ and $\mathcal{I}_2$ can be done using the following identities for Bessel functions
\begin{align}
	J_{0}\left(x\right) ~ &\coloneqq ~ \dfrac{2}{\pi} \int_{0}^{\infty} \mathrm{d} y \, \sin \left( x \cosh y \right) \quad \text{if} \quad x > 0 \, , \\
	Y_{0}\left(x\right) ~ &\coloneqq ~ -\dfrac{2}{\pi} \int_{0}^{\infty} \mathrm{d} y \, \cos \left( x \cosh y \right) \, , \\
	K_{0}\left(x\right) ~ &\coloneqq ~ \int_{0}^{\infty} \mathrm{d} y \, \cos \left( x \sinh y \right) \, ,
\end{align}
to find
\begin{align}
    \mathcal{I}_1 ~ = ~ 2 K_0\left(m r\right) \quad \text{and}
 \quad \mathcal{I}_2 ~ = ~ 0 \, .
\end{align}
Using this expression, the correlator can be written in a general Lorentz-invariant form as
\begin{align}\label{eqn:phiphiMassiveSpacelike}
    \bva \phi\left(x_1\right) \phi\left(x_2\right) \kva ~ &= ~ \dfrac{-1}{4 \pi^2 r} \dfrac{\partial}{\partial r} \left(K_0\left(m \sqrt{r^2 - t^2}\right)\right) ~ \text{when} ~ \left(x_1 - x_2\right)^2 = r^2 - t^2 > 0 \nonumber \\
    &= ~ \dfrac{m}{4 \pi^2 r} K_1\left(m r\right) \quad \text{when} \quad t ~ = ~ 0 \, .
\end{align}
It is straightforward to check that the massless limit of this correlator agrees with the expression we have in \eqref{eqn:phiphiBulk}. Moreover, it is now evident that
\begin{align}
    \bva \Big[\phi\left(x_1\right) , \phi\left(x_2\right)\Big] \kva ~ = ~ 0 \qquad \text{when} \qquad \left(x_1 - x_2\right)^2 = r^2 - t^2 > 0 \, .
\end{align}
Noting that $t = x^0_2 - x^0_1$ and taking a time derivative of \eqref{eqn:phiphiMassiveSpacelike} with respect to $x^0_2$, we find, when $r^2 - t^2 > 0$, that
\begin{align}\label{eqn:massivephipi}
    \bva \phi\left(x_1\right) \pi\left(x_2\right) \kva ~ = ~ \dfrac{-1}{4 \pi^2 r} \dfrac{\partial}{\partial r} \left(\dfrac{m t K_1\left(m \sqrt{r^2 - t^2}\right)}{\sqrt{r^2 - t^2}}\right) \, .
\end{align}
Taking a further derivative with respect to $x^0_1$, we now find 
\begin{align}\label{eqn:massivepipi}
    \bva \pi\left(x_1\right) \pi\left(x_2\right) \kva ~ &= ~ \dfrac{1}{4 \pi^2 r} \dfrac{\partial}{\partial r} \left(\dfrac{m^2 t^2 K_0\left(m \sqrt{r^2 - t^2}\right)}{r^2 - t^2} + \dfrac{m \left(r^2 + t^2\right) K_1\left(m \sqrt{r^2 - t^2}\right)}{\sqrt{\left(r^2 - t^2\right)^3}}\right) \nonumber \\
    &= ~ - \dfrac{m^2 K_2\left(m r\right)}{4 \pi^2 r^2} \qquad \text{when} \qquad t = 0 \, .
\end{align}

\paragraph{Time-like separation: } In this case, we have that $\left(x_1 - x_2\right)^2 = \left(- t^2 + r^2\right) < 0$. Choosing a frame where $r=0$, the integrals $\mathcal{I}_{1}$ and $\mathcal{I}_2$ can be done in similar fashion as before to find
\begin{align}
    \mathcal{I}_1 ~ = ~ - \pi Y_0\left(m t\right) \quad \text{and} \quad \mathcal{I}_2 ~ = ~ \pi \text{sgn}\left(t\right) J_0\left(m \abs{t}\right) \, .
\end{align}
As before, these two can be combined and written in a Lorentz-invariant form as
\begin{align}\label{eqn:phiphiMassiveTimelike}
    \bva \phi\left(x_1\right) \phi\left(x_2\right) \kva ~ &= ~ \dfrac{i}{8 \pi^2 r} \dfrac{\partial}{\partial r} \dfrac{4}{\pi} \Big( \text{sgn}\left(x^0_1 - x^0_2\right) J_0\left(m \sqrt{t^2 - r^2}\right) - i Y_0\left(m \sqrt{t^2 - r^2} \right) \Big) \nonumber \\ 
    &\hspace{3cm} \text{when} \quad \left(x_1 - x_2\right)^2 = r^2 - t^2 < 0 \, .
\end{align}
Therefore, for the commutator, we have that
\begin{align}\label{eqn:massivephiphiComm-timelike}
    \bva \Big[\phi\left(x_1\right) , \phi\left(x_2\right)\Big] \kva ~ &= ~ \dfrac{i}{4 \pi^2 r} \dfrac{\partial}{\partial r} \dfrac{4}{\pi} \Big(\text{sgn}\left(x^0_1 - x^0_2\right) J_0\left(m \sqrt{t^2 - r^2}\right) \Big) \nonumber \\
    &\hspace{3cm} \text{when} \quad \left(x_1 - x_2\right)^2 = r^2 - t^2 < 0 \, .
\end{align}

\section{Inversion for a correlator from its moments}\label{app:inversion}
In this appendix, we describe what information we may infer about the functions $f_p\left(\left\{\vec{x}_p\right\}\right)$ given measurements of the kind discussed in \secref{sec:bulkcorrelators} and \secref{sec:boundarycorrelators}.

\subsection{Inversion for bulk correlators}\label{app:bulkinversion}
We begin with the result for the generic bulk correlator of interest \eqref{eqn:bulk-correlators}:
\begin{align}
    &\bpsi \pi\left(\vec{y}\right) \phi^m\left(\left\{\vec{y}_m\right\}\right) \kpsi - \bva \pi\left(\vec{y}\right) \phi^m\left(\left\{\vec{y}_m\right\}\right) \kva ~ = ~ \nonumber \\
    &\quad = - i \gamma \int \sum_{i = 1}^\infty \left(\prod_{j=1}^{i} \mathrm{d}^3 \vec{x}_k\right) f_i\left(\left\{\vec{x}_i\right\}\right) \sum_{k=1}^i \bva :\phi^{i-1}\left(\left\{\vec{x}_{i-1}\right\}\right): \phi^m\left(\left\{\vec{y}_m\right\}\right) \kva \left[i \delta^3\left(\abs{\vec{x}_k - \vec{y}}\right)\right] \, .
\end{align}
As we saw in \secref{sec:bulkonept}, the one-point function allows us to determine the function $f_1\left(\vec{x}_1\right)$ defining a one-particle state. To illustrate how we may invert this integral equation for the function $f_i\left(\left\{\vec{x}_i\right\}\right)$ in general, let us now specialise to the case of the two-point function where $m=1$ as in \eqref{eqn:bulktwoptfn}: 
\begin{align}\label{eqn:O2kernel}
    \mathcal{O}_2\left(\vec{y}_1 , \vec{y}_2\right) ~ &\coloneqq ~ \bpsi \pi\left(\vec{y}_1\right) \phi\left(\vec{y}_2\right) \kpsi - \bva \pi\left(\vec{y}_1\right) \phi\left(\vec{y}_2\right) \kva \nonumber \\
    &= ~ \dfrac{m \gamma}{4 \pi^2} \int \mathrm{d}^3 \vec{x}_1 \left(\dfrac{K_1\left(m \left|\vec{y}_2 - \vec{x}_1\right|\right) }{ \left|\vec{y}_2 - \vec{x}_1\right|}\right) \Big[f_2\left(\vec{x}_1 , \vec{y}_1\right) + f_2\left(\vec{y}_1 , \vec{x}_1\right)\Big] \nonumber \\
    &\eqqcolon ~ \dfrac{m \gamma}{4 \pi^2} \int \mathrm{d}^3 \vec{x}_1 \left(\dfrac{K_1\left(m \left|\vec{y}_2 - \vec{x}_1\right|\right) }{ \left|\vec{y}_2 - \vec{x}_1\right|}\right) \tilde{f}_2\left(\vec{y}_1 , \vec{x}_1 \right) \, .
\end{align}
We first observe that the dependence on one of the arguments, $\vec{y}_1$ is explicit in this equation. Therefore, given a fixed $\vec{y}_2$, we may simply make various measurements everywhere inside the ball to determine the dependence of the function $\tilde{f}_2$ on $\vec{y}_1$. Moreover, in order for this correlator to be well-behaved, we assume that the function $f_2\left(\vec{x}_1 , \vec{y}_1\right)$ is such that the correlator does not blow up when $\vec{x}_1$ approaches $\vec{y}_2$. In what follows, we will use the identities:
\begin{align}
    &\dfrac{1}{\Gamma\left(n + 1\right)} \int_0^\infty \mathrm{d}s \, s^n \, e^{- s \left|\vec{y}_2 - \vec{x}_1\right|^m} ~ = ~ \dfrac{1}{\left|\vec{y}_2 - \vec{x}_1\right|^{m\left(n+1\right)}} \\ 
    &K_n\left(z\right) ~ = ~ \Gamma\left[n + \dfrac{1}{2}\right] \dfrac{\left(2 z\right)^n}{\sqrt{\pi}} \int_0^\infty \mathrm{d} u \dfrac{\cos\left(u\right)}{\left(u^2 + z^2\right)^{n + \frac{1}{2}}} \, ,
\end{align}
combining which we have
\begin{align}\label{eqn:BesselKoverr}
    \dfrac{K_q\left(m \left|\vec{y}_2 - \vec{x}_1\right|\right)}{\left|\vec{y}_2 - \vec{x}_1\right|^l} ~ &= ~ \Gamma\left[q + \dfrac{1}{2}\right] \dfrac{\left(2 m\right)^q}{\sqrt{\pi}} \left|\vec{y}_2 - \vec{x}_1\right|^{q - l} \int_0^\infty \mathrm{d} u \dfrac{\cos\left(u\right)}{\left(u^2 + m^2 \left|\vec{y}_2 - \vec{x}_1\right|^2\right)^{q + \frac{1}{2}}} \nonumber \\
    &= ~ \dfrac{\left(2 m\right)^q}{\sqrt{\pi}} \left|\vec{y}_2 - \vec{x}_1\right|^{q - l} \nonumber \\
    &\hspace{0.5cm} \times \int_0^\infty \mathrm{d} u \cos\left(u\right) \int_0^\infty \mathrm{d}s \, s^{q - \frac{1}{2}} \exp\left[- s \left(u^2 + m^2 \left|\vec{y}_2 - \vec{x}_1\right|^2\right)\right] \, .
\end{align}
Therefore, we may write
\begin{align}\label{eqn:O2}
    \mathcal{O}_2\left(\vec{y}_1 , \vec{y}_2\right) ~ &= ~ \dfrac{m^2 \gamma}{2 \pi^2 \sqrt{\pi}} \int \mathrm{d}^3 \vec{x} \Bigg[\int_0^\infty \mathrm{d} u \cos\left(u\right) \nonumber \\
    &\hspace{3cm} \times \int_0^\infty \mathrm{d}s \, s^{\frac{1}{2}} \exp\left[- s \left(u^2 + m^2 \left|\vec{y}_2 - \vec{x}\right|^2\right)\right]\Bigg] \tilde{f}_2\left(\vec{y}_1 , \vec{x}\right) \nonumber \\
    &= ~ \dfrac{m^2 \gamma}{2 \pi^2 \sqrt{\pi}} \sum_{\ell m} \int_0^\infty \left|\vec{x}\right|^2 \mathrm{d} \left|\vec{x}\right| \Bigg[\int_0^\infty \mathrm{d} u \cos\left(u\right) \int_0^\infty \mathrm{d}s \, s^{\frac{1}{2}} e^{- s \left(u^2 + m^2 \left|\vec{y}_2\right|^2 + m^2 \left|\vec{x}\right|^2\right)} \nonumber \\
    &\hspace{5.5cm} \int \mathrm{d} \Omega^{(2)}_{\vec{x}} e^{2 s m^2 \vec{y}_2 \cdot \vec{x} }\Bigg] \tilde{f}^{\ell m}_2\left(\vec{y}_1 , \left|\vec{x}\right|\right) Y_\ell^m \left(\Omega^{(2)}_{\vec{x}}\right) \nonumber \\
    &= ~ \dfrac{2 m^2 \gamma}{\pi \sqrt{\pi}} \sum_{\ell m} \int_0^\infty \left|\vec{x}\right|^2 \mathrm{d} \left|\vec{x}\right| \Bigg[\int_0^\infty \mathrm{d} u \cos\left(u\right) \int_0^\infty \mathrm{d}s \, s^{\frac{1}{2}} e^{- s \left(u^2 + m^2 \left|\vec{y}_2\right|^2 + m^2 \left|\vec{x}\right|^2\right)} \nonumber \\
    &\hspace{2.5cm} \left(2 s m^2 \left|\vec{y}_2\right|\right)^\ell i_\ell \left(2 s m^2 \left|\vec{y}_2\right| \left|\vec{x}\right| \right) \Bigg] \tilde{f}^{\ell m}_2\left(\vec{y}_1 , \left|\vec{x}\right|\right) Y_{\ell}^m \left(\Omega_{\vec{y}_2}\right) \, ,
\end{align}
where $i_\ell \left(x\right)$ is the spherical Bessel function that respects the identity
\begin{align}
    \int \mathrm{d} \Omega^{(2)}_{\vec{x}} e^{2 s m^2 \vec{y}_2 \cdot \vec{x}} Y_{\ell}^m \left(\Omega\right) ~ = ~ 4 \pi \left(2 s m^2 \left|\vec{y}_2\right|\right)^\ell Y_{\ell}^m \left(\Omega_{\vec{y}_2}\right) i_\ell \left(2 s m^2 \left|\vec{y}_2\right| \left|\vec{x}\right| \right) \, .
\end{align}
The spherical Bessel I function can also be written in terms of reverse Bessel functions as
\begin{align}
    i_\ell \left(x\right) ~ = ~ \dfrac{e^{- x} \theta_\ell \left(x\right) - e^x \theta_\ell \left(-x\right)}{2 \left(- x\right)^{\ell + 1}} \qquad \text{with} \qquad \theta_\ell \left(x\right) ~ \coloneqq ~ \sum_{n = 0}^\ell \dfrac{x^{\ell - n}}{2^n n!} \dfrac{\left(\ell + n\right)!}{\left(\ell - n\right)!} \, .
\end{align}
This implies that
\begin{align}
    i_\ell \left(2 s m^2 \left|\vec{y}_2\right| \left|\vec{x}\right| \right) ~ &= ~ \sum_{n = 0}^\ell \left[\dfrac{e^{- \left(2 s m^2 \left|\vec{y}_2\right| \left|\vec{x}\right|\right)} - e^{\left(2 s m^2 \left|\vec{y}_2\right| \left|\vec{x}\right|\right)} \left(-1\right)^{\ell - n}}{\left(2 s m^2 \left|\vec{y}_2\right| \left|\vec{x}\right|\right)^{n + 1}} \right] \dfrac{\left(\ell + n\right)!}{2^{n+1} n! \left(-1\right)^{\ell + 1} \left(\ell - n\right)!} \, .
\end{align}
Therefore, we have that
\begin{align}\label{eqn:s-integral}
    &\int_0^\infty \mathrm{d} s \, s^{\frac{1}{2}} e^{- s \left(u^2 + m^2 \left|\vec{y}_2\right|^2 + m^2 \left|\vec{x}\right|^2\right)} \left(2 s m^2 \left|\vec{y}_2\right|\right)^\ell i_\ell \left(2 s m^2 \left|\vec{y}_2\right| \left|\vec{x}\right|\right) ~ = ~ \nonumber \\
    &\qquad = ~ \sum_{n=0}^\ell \Bigg(\dfrac{(-1)^{-n} \left(u^2 + m^2 \left(\left|\vec{x}\right| - \left|\vec{y}_2\right|\right)^2\right)^{-\ell +n-\frac{1}{2}}-(-1)^{-\ell } \left(u^2 + m^2 \left(\left|\vec{x}\right| + \left|\vec{y}_2\right|\right)^2\right)^{-\ell + n - \frac{1}{2}}}{2^{2n + 2 - \ell} m^{2n + 2 - 2\ell} \left|\vec{y}_2\right|^{-\ell} \left(\left|\vec{x}\right|\left|\vec{y}_2\right|\right)^{n+1}} \nonumber \\
    &\hspace{4cm} \times \Gamma\left[\ell - n + \dfrac{1}{2}\right] \dfrac{\left(\ell + n\right)!}{n! \left(\ell - n\right)!} \Bigg) \qquad \text{with} \quad n ~ < ~ \dfrac{1}{2} + \ell \, .
\end{align}
We may now easily carry out the $u$-integral in \eqref{eqn:O2} to find:
\begin{align}
    &\int_0^\infty \mathrm{d}u \cos\left(u\right) \int_0^\infty \mathrm{d} s \, s^{\frac{1}{2}} e^{- s \left(u^2 + m^2 \left|\vec{y}_2\right|^2 + \left|\vec{x}\right|^2\right)} \left(2 s m^2 \left|\vec{y}_2\right|\right)^\ell i_\ell \left(2 s m^2 \left|\vec{y}_2\right| \left|\vec{x}\right|\right) ~ = ~ \nonumber \\
    &~ = ~ \sum_{n=0}^\ell \dfrac{\sqrt{\pi } m^{\ell - n} \left(\ell + n\right)!}{2^{n+2} m^2 n! \left(\ell - n\right)!} \dfrac{\left|\vec{y}_2\right|^{\ell}}{\left|\vec{x}\right|^{n+1} \left|\vec{y}_2\right|^{n+1}} \nonumber \\
    &\hspace{2cm} \times \Bigg[(-1)^{-n} \dfrac{K_{\ell - n } \left( m \left| \left|\vec{x}\right| - \left|\vec{y}_2\right| \right| \right)}{\left| \left|\vec{x}\right| - \left|\vec{y}_2\right| \right|^{\ell - n}} - (-1)^{-\ell} \dfrac{K_{\ell - n } \left( m \left|
    \vec{x}\right| + \left|\vec{y}_2\right| \right)}{\left|\vec{x}\right| + \left|\vec{y}_2\right|^{\ell - n}} \Bigg] \, .
\end{align}
By making many measurements at various angular positions of $\vec{y}_2$, we may expand them in spherical harmonics. Then, from the orthogonality of the spherical harmonics, \eqref{eqn:O2} implies
\begin{align}\label{eqn:inversionbulkf2}
    \mathcal{O}^{\ell m}_2 \left(\vec{y}_1 , \left|\vec{y}_2\right|\right) ~ &= ~ \dfrac{\gamma}{\pi} \sum_{n=0}^\ell \int_0^\infty \dfrac{\left|\vec{x}\right|^2 \mathrm{d} \left|\vec{x}\right| \left|\vec{y}_2\right|^{\ell}}{\left|\vec{x}\right|^{n+1} \left|\vec{y}_2\right|^{n+1}} \dfrac{m^{\ell - n} \left(\ell + n\right)!}{2^{n+1} n! \left(\ell - n\right)!} \tilde{f}^{\ell m}_2\left(\vec{y}_1 , \left|\vec{x}\right|\right) \nonumber \\ 
    &\quad \times \Bigg[(-1)^{-n} \dfrac{K_{\ell - n } \left( m \left| \left|\vec{x}\right| - \left|\vec{y}_2\right| \right| \right)}{ \left| \left|\vec{x}\right| - \left|\vec{y}_2\right| \right|^{\ell - n}} - (-1)^{-\ell} \dfrac{K_{\ell - n } \left( m \left| \vec{x}\right| + \left|\vec{y}_2\right| \right)}{ \left|\left|\vec{x}\right| + \left|\vec{y}_2\right|\right|^{\ell - n}} \Bigg] \, .
\end{align}
This integral can now be split into a region where $\left|\vec{x}\right| < \left|\vec{y}_2\right|$ and another where $\left|\vec{x}\right| > \left|\vec{y}_2\right|$. In the former, we may expand the integrand in a power series in $\frac{\left|\vec{x}\right|}{\left|\vec{y}_2\right|}$ whereas in the latter, the expansion may be done in $\frac{\left|\vec{y}_2\right|}{\left|\vec{x}\right|}$. This expansion takes the following form:
\begin{align}\label{eqn:O2expansion}
    \mathcal{O}^{\ell m}_2 \left(\vec{y}_1 , \left|\vec{y}_2\right|\right) ~ &= ~ \dfrac{\gamma}{\pi} \sum_{n=0}^\ell \int_0^\infty \dfrac{\left|\vec{x}\right|^2 \mathrm{d} \left|\vec{x}\right| \left|\vec{y}_2\right|^{\ell}}{\left|\vec{x}\right|^{n+1} \left|\vec{y}_2\right|^{n+1}} \dfrac{m^{\ell - n} \left(\ell + n\right)!}{2^{n+1} n! \left(\ell - n\right)!} \tilde{f}^{\ell m}_2\left(\vec{y}_1 , \left|\vec{x}\right|\right) \nonumber \\ 
    &\hspace{1cm} \times \begin{cases*}
    \sum_{q=0}^\infty \left|\vec{x}\right|^q c^{(q)}_{\ell n} \left(\left|\vec{y}_2\right|\right) \qquad \text{when} \qquad \dfrac{\left|\vec{x}\right|}{\left|\vec{y}_2\right|} ~ < ~ 1 \\
    \sum_{q=0}^\infty \left|\vec{y}_2\right|^q c^{(q)}_{\ell n} \left(\left|\vec{x}\right|\right) \qquad \text{when} \qquad \dfrac{\left|\vec{x}\right|}{\left|\vec{y}_2\right|} ~ > ~ 1 \, ,
    \end{cases*}
\end{align}
where the kernels can be worked out systematically
\begin{align}
    c^{(0)}_{\ell n} \left(r\right) ~ &= ~ \left[\left(-1\right)^n - \left(-1\right)^\ell\right] \left(\dfrac{1}{r}\right)^{\ell - n} K_{\ell - n} \left(m r\right) \nonumber \\
    c^{(1)}_{\ell n} \left(r\right) ~ &= ~ \left[\left(-1\right)^n + \left(-1\right)^\ell\right] m \left(\dfrac{1}{r}\right)^{\ell - n} K_{\ell - n + 1} \left(m r\right) \nonumber \\
    c^{(2)}_{\ell n} \left(r\right) ~ &= ~ \dfrac{1}{2} \left[\left(-1\right)^\ell - \left(-1\right)^n\right] m \left(\dfrac{1}{r}\right)^{\ell - n + 1} \Big[m r K_{\ell - n} \left(m r\right) + \left(2 \ell - 2 n + 1\right) K_{\ell - n + 1}\left(m r\right)\Big] \nonumber \\
    \vdots \nonumber
\end{align}
This expansion results in a system of linear equations for different points of measurements $\left|\vec{y}_2\right|$ with coefficients being the various moments of the function $\tilde{f}_2$ that we aim to determine. At this stage, massive and massless cases need to be treated separately. 

It is rather subtle to take the massless limit in \eqref{eqn:inversionbulkf2} because in that case, the $u$-integral results in divergent contributions in the $\ell = n$ sector. The cleanest way to address this is to simply start with the massless kernel in \eqref{eqn:O2kernel}. Then, we may perform the spherical averaging of the exponential to obtain the spherical Bessel function $i_\ell$. Thereafter, we may separate out each partial wave independently to split it into internal and external moments. Alternatively, we may write the result of the spherical average in the form of reverse Bessel polynomials as we did in the massive case. In the latter case, however, the $\ell = n$ sector has to be treated separately before performing the $s$ and $u$ integrals. In that case, it turns out that the expansion in the region $\left|\vec{x}\right| > \left|\vec{y}_2\right|$ comes with negative powers of $\left|\vec{x}\right|$ instead of the cominbations of Bessel functions as we found above. Then, a simple redefinition as $\left|\vec{x}\right| = \frac{1}{w}$ turns them into positive moments in terms of the new parameter $w$. Then we may split the integral into two parts from $0$ to $\left|\vec{y}_2\right|$ and then from $\left|\vec{y}_2\right|$ to $\infty$. This way, solving the system of linear equations for various measurements at various radial positions $\left|\vec{y}_2\right|$ allows us to determine the dependence of the function everywhere in space, in terms of its internal (within the ball) and external (outside the ball) moments. Clearly, this procedure works for all functions that have a well-defined inverse Laplace transform which is what we demand of the functions defining the states. The conclusion is then that we have knowledge of the function $\tilde{f}_2$ everywhere in space on one argument whereas its knowledge of the other argument is limited to within the ball.

In the massive case, however, the situation is slightly more complicated. The internal moments are identical and we may determine the function inside the ball entirely using the same method. However, the expansion outside the ball contains kernels which are combinations of the modified Bessel functions as can be seen in \eqref{eqn:O2expansion}. This makes it harder to determine exactly what information about the function $\tilde{f}_2$ we have access to outside the ball.

\paragraph{Basis of functions:} To address this concern of external moments in the case massive fields, we may approach the problem in an alternative fashion. We may pick an appropriate basis of functions to expand $\tilde{f}^{\ell m}_2\left(\vec{y}_1, \left|\vec{x}\right|\right)$ in. This gives us undetermined coefficients that govern the linear combination of the basis functions that yield our function of interest $\tilde{f}^{\ell m}_2\left(\vec{y}_1 , \left|\vec{x}\right|\right)$. Now, by making many different measurements for each value of $\left|\vec{y}_2\right|$, we may determine these coefficients by solving a system of linear equations. Therefore, by measuring the correlator for all values of $\left|\vec{y}_2\right|$ in a small region (which is still an infinite number of measurements), we may determine the function exactly. A finite number of measurements yields the function within a certain error that depends on various factors such as appropriateness of the choice of basis functions, etc. This of course requires us to have knowledge of a good basis to begin with, that is practical and appropriate for the physical system at hand. Moreover, the basis functions must be analytic everywhere in space to prevent the coefficients of the basis expansion from being dependent on space. This implies that sufficiently precise measurements inside the ball will give us all information about the function in both arguments everywhere in space. However, this is an artefact of our prior knowledge of a good basis and the approximation that the function defining the state $f_2$ can be expanded in a linear combination of analytic functions.\footnote{Generically, the functions defining the state were not assumed to be analytic. However, this expansion into a basis approximates the functions in a superposition of analytic basis elements.} This is a counter-intuitive result as we expect that correlators inside the ball should have no access to unitary local excitations far away from the ball. In a practical sense, this is still true in that given some experimental precision, two states may very well differ dramatically far outside the ball provided that their difference within the ball falls within that experimental precision that the low energy observer has no access to. 

On the other hand, this method of expanding into a basis of functions can be done for both the massive and massless cases alike. Notwithstanding this, there is still an important difference between the two cases in practice. Since the kernels in the massive case are modified Bessel functions outside the ball, they fall off exponentially. Whereas the massless case exhibits power-law fall-offs. This means that extracting the coefficients corresponding to modes beyond a Compton wavelength (of the massive field) away from the ball is increasingly harder. Whereas the massless case has long range effects that are easier to detect.\footnote{We are grateful to R. Loganayagam for discussions on this point and several other subtleties discussed in this appendix.} 

In the following sub-section, we will see that when there are multiple variables in the function to be inverted for, there is a general multi-variate moment expansion from which we will be able to invert for the corresponding function $f_n$ where one of its arguments is restricted to be inside the sphere of radius $R$.

\subsection{Inversion for boundary correlators}\label{app:boundaryinversion}
In the main text, we found the boundary correlator that was relevant for the bulk one-point function to be
\begin{align}
    \mathcal{O}_1\left(\vec{z}\right) ~ &\coloneqq ~ \bpsi H_R \pi\left(\vec{z}\right) \kpsi - \bva \pi\left(\vec{z}\right) \kva \nonumber \\
    &= ~ \dfrac{- m^2 \gamma}{4 \pi^2} \int \mathrm{d}^3 \vec{x} f_1\left(\vec{x}\right) \dfrac{K_2\left(m \left|\vec{x} - \vec{z}\right|\right)}{\left|\vec{x} - \vec{z}\right|^2} \Theta\left( R - \abs{\vec{x}}\right) \, .
\end{align}
Now, using \eqref{eqn:BesselKoverr} and the following steps, we may write this as
\begin{align}\label{eqn:inversionBulkonepointf1}
    \mathcal{O}_1\left(\vec{z}\right) &= \dfrac{- m^4 \gamma}{\pi^2 \sqrt{\pi}} \int \mathrm{d}^3 \vec{x} f_1\left(\vec{x}\right) \int_0^\infty \mathrm{d} u \cos\left(u\right) \int_0^\infty \mathrm{d}s \, s^{\frac{3}{2}} e^{- s \left(u^2 + \left|\vec{y}_2 - \vec{x}_1\right|^2\right)} \Theta\left( R - \abs{\vec{x}}\right) \, .
\end{align}
The only difference between the $s$-integral in this expression and the corresponding integral in \eqref{eqn:s-integral} is an extra power of $s$ in front of the exponential. Therefore, while the resulting expression after carrying out the $s$ and $u$ integrals, the conclusion can explicitly be checked to be the same that we may invert the above correlator for the function $f_1\left(\vec{x}\right)$ using all its moments. As is evident, the explicit appearance of the $\Theta$-function restricts our knowledge of the one-particle exponent to within the sphere of radius $R$.

Similarly, the boundary correlator that corresponds to the bulk two-point function is
\begin{align}
    \mathcal{O}\left(\vec{z} , \vec{z}_1\right) ~ &\coloneqq ~ \bpsi H_R \pi\left(\vec{z}\right) \phi\left(\vec{z}_1\right) \kpsi - \bva H_R \pi\left(\vec{z}\right) \phi\left(\left\{\vec{z}_m\right\}\right) \kva \nonumber \\
    &= ~ \dfrac{- m^3 \gamma}{16 \pi^4} \int \mathrm{d}^3 \vec{x}_1 \mathrm{d}^3 \vec{x}_2 \tilde{f}_2\left(\vec{x}_1 , \vec{x}_2\right) \Bigg[ \dfrac{K_2\left(m \left|\vec{x}_2 - \vec{z}\right|\right)}{\left|\vec{x}_2 - \vec{z}\right|^2} \dfrac{K_1\left(m \left|\vec{x}_1 - \vec{z}_1\right|\right)}{\left|\vec{x}_1 - \vec{z}_1\right|} \Theta\left( R - \abs{\vec{x}_1}\right) \Bigg] \, .
\end{align}
This integral contains two fractions, both of which have appeared earlier in this appendix resulting in a product of expressions of the kind \eqref{eqn:BesselKoverr}. Therefore, we may proceed as before to carry out the integrals arising from this product. Furthermore, we may expand the left hand side of the correlator in partial waves by moving around both along the angular directions of $\vec{z}$ and $\vec{z}_1$. Consequently, we will find internal and external moments as in the bulk. Therefore, from this bi-variate moment expansion, we may invert for the symmetrised function $\tilde{f}_2\left(\vec{x}_1 , \vec{x}_2\right)$ everywhere provided one of the arguments (labelled $\vec{x}_1$ in this case) is inside the sphere of radius $R$. The dependence on the other argument is identical to the bulk correlator because the kernel is identical. Therefore, without further inverting the functions, we already see that all information that we have of the function $f_2$ from the bulk correlators is also available to the boundary correlator in a small radial band. Of course, we may proceed to invert for the function either in its moments or alternatively, again, by expanding the function $\tilde{f}_2$ in an appropriate basis to turn the problem into a system of linear equations. This is precisely the same information as contained in the bulk two-point function \eqref{eqn:bulktwoptfn}.

All higher-point functions will similarly involve multi-variate moment expansions or systems of linear equations arising from expanding them into a basis, from which we will be able to invert for all information about the corresponding symmetrised support function $\tilde{f}_n$ where one of the arguments will be restricted to be inside the sphere of radius of $R$.

\printbibliography

\end{document}